\documentclass{aastex62}
\usepackage{amsmath}

\usepackage{mathtools, nccmath}

\graphicspath{{./}{figures/}}

\accepted{October 6, 2020}
\submitjournal{AJ}

\shorttitle{The End of Galaxy Surveys}
\shortauthors{Rhodes et al.}
\begin{document}

\title{The End of Galaxy Surveys}

\correspondingauthor{Jason Rhodes}
\email{jason.d.rhodes@jpl.nasa,gov}

\author{Jason D. Rhodes}
\affil{Jet Propulsion Laboratory, California Institute of Technology \\
4800 Oak Grove Drive \\
Pasadena, CA 91109, USA}
\affiliation{Kavli Institute for the Physics and Mathematics of the Universe (WPI), University of Tokyo, Kashiwa 277-8583, Japan}

\author{Eric Huff}
\affiliation{Jet Propulsion Laboratory, California Institute of Technology \\
4800 Oak Grove Drive \\
Pasadena, CA 91109, USA}

\author{Daniel Masters}
\affiliation{Jet Propulsion Laboratory, California Institute of Technology \\
4800 Oak Grove Drive \\
Pasadena, CA 91109, USA}
\affiliation{NASA Postdoctoral Program (NPP) Fellow}
\affiliation{IPAC, Caltech, 1200 E. California Blvd., Pasadena, CA 91125, USA}

\author{Anna Nierenberg}
\affiliation{Jet Propulsion Laboratory, California Institute of Technology \\
4800 Oak Grove Drive \\
Pasadena, CA 91109, USA}
\affiliation{NASA Postdoctoral Program (NPP) Fellow}

%https://academic.oup.com/mnras/article/479/4/5184/5050078



\begin{abstract}
For nearly a century, imaging and spectroscopic surveys of galaxies have given us information about the contents of the universe.  We attempt to define the logical endpoint of such surveys by defining not the next galaxy survey, but the final galaxy survey at NIR wavelengths; this would be the galaxy survey that exhausts the information content useful for addressing extant questions. Such a survey would require incredible advances in a number of technologies  and the survey details will depend on the as yet poorly constrained properties of the earliest galaxies.  Using an exposure time calculator, we define nominal surveys for extracting the useful information for three science cases: dark energy cosmology, galaxy evolution, and supernovae. We define scaling relations that trade off sky background, telescope aperture, and focal plane size to allow for a survey of a given depth over a given area.  For optimistic assumptions, a  280m telescope with a marginally resolved focal plane of 20 deg$^2$ operating at L2  could potentially exhaust the cosmological  information content of galaxies in a 10 year survey.  For galaxy evolution (making use of gravitational lensing to magnify the earliest galaxies) and SN, the same telescope would suffice. We discuss the technological advances needed to complete the last galaxy survey. While the final galaxy survey remains well outside of our technical reach today, we present scaling relations that show how we can progress toward the goal of exhausting the information content encoded in the shapes, positions, and colors of  galaxies.

\end{abstract}

\keywords{cosmology-- galaxy evolution --- surveys}


\section{Introduction} \label{sec:intro}

In the coming decade, optical and near infrared (NIR) sky surveys will reach increasing depths over much of the sky. Euclid \citep{2011arXiv1110.3193L}, the Nancy Grace Roman Space Telescope  \citep[hereafter Roman and formerly known as WFIRST;][]{2019arXiv190205569A}, the Legacy Survey of Space and Time to be taken with the Vera Rubin Observatory \citep[LSST;][]{2009-Book-LSST}, and the Dark Energy Spectroscopic Instrument \citep[DESI;][]{2016arXiv161100036D}, among others, will take images and spectra over nearly the entire extragalactic sky  to unprecedented depths and resolution, allowing for transformative science in cosmology and galaxy evolution. However, while the observable universe is almost unimaginably large, it is finite. 
There is therefore a finite amount of information about the universe encoded in the positions, colors, and shapes of galaxies. 
In this paper we propose a series of surveys that would exhaust the information content in galaxies for the study of dark energy cosmology and observe the earliest SN in the universe. Furthermore we consider what photometric and spatially unresolved spectroscopic observations could address fundamental questions about galaxy formation and evolution for the faintest galaxies in the Universe.
We do not seek to justify that these proposed future surveys are necessary or even advisable goals for the near future; rather, we seek to show what surveys would bring the current era of galaxy surveys to its logical end point. 

This paper is organized as follows. 
In \S\ref{sec:telparams} we provide a pedagogical review of telescope, instrument, and other observing parameters which determine the detectability of an object. In \S\ref{sec:surveys} we lay out three overarching science drivers which, if pursued,  would complete our mapping of galaxies and supernovae for specific science goals related to cosmology and galaxy evolution. This discussion leads to order of magnitude estimates in observatories parameters needed to  complete these surveys.
%In \S\ref{sec:info} we define metrics for the information content available in the optical and near infrared images and (spatially unresolved) spectra of galaxies in the visible universe for cosmology, galaxy evolution, and supernovae. 
%\S\ref{sec:surveys} describe the surveys needed to exhaust this information content.  
%[\textbf{Anna, this new version then doesn't really talk about information %content, which I felt was the primary focus of the paper. Have I missed %something about your ideas for the new organization?}]
In  \S\ref{sec:mission} we describe  the technological tall poles to achieving and end to galaxy surveys.  
Finally, in  \S\ref{sec:conclusions} we offer some concluding remarks.

\section{Observing Parameters} \label{sec:telparams}
In this section we give a pedagogical review of the parameters that determine the achievable depth of a given survey. We  first discuss photometry and the talk about the added time that would be needed to get spectroscopy at different levels of accuracy. 

\subsection{Imaging} \label{sec:img}
The speed of a survey of given area and depth depends on the field of view (FOV) of the instrument(s) together with the integration time required to achieve the desired signal-to-noise ratio (SNR) on the faintest sources. 
To determine the time it would take to conduct a survey to a given point source depth, we can invert the standard SNR equation:

\begin{equation} \label{snr}
    \mathrm{SNR} = 
    \frac{S_{o}Qt} 
    {\sqrt{S_{o}Qt+S_{s}Qtn_{p}+S_{d}tn_{p}+R^{2}n_{p}}},
\end{equation}

\noindent where $S_{o}$ is the source flux (photon/sec), $Q$ is the quantum efficiency of the detector, t is the exposure time in seconds, $S_{s}$ is the sky background in photon/sec/pixel, $S_{d}$ is the dark current in electron/sec/pixel, $R$ is the read noise in electron/pixel, and $n_{p}$ is the number of pixels over which the measurement is made. See Table~\ref{tab:quantities} for a description of the relevant parameters. 
Solving for $t$, we have, for perfect observing efficiency (i.e. $\epsilon=1$):
\begin{equation} \label{inverse_snr}
    t = \frac{1}{2} \left( \frac{\mathrm{SNR}}{S_{o}Q} \right) ^{2} \left(S_{o}Q + S_{s}Qn_{p}+S_{d}n_{p} \right) + \frac{1}{2}\sqrt{ \left( \left( \frac{\mathrm{SNR}}{S_{o}Q} \right)^{2} \left(S_{o}Q + S_{s}Qn_{p}+S_{d}n_{p} \right) \right)^{2} + 4n_{p} \left(\frac{\mathrm{SNR}\times R}{S_{o}Q}\right)^{2} } 
\end{equation}

\noindent This expression can further be written in terms of the telescope diameter, $D$. Both $S_{s}$ and $S_{o}$ scale as $D^{2}$, while the number of pixels $n_{p}$ an object covers scales as $D^{-2}$ for a   diffraction limited observatory. This means that the sky background near a point source is invariant as the telescope gets larger, and the time to reach a given point source depth goes as $D^{-4}$. 
  However, this is only true for unresolved sources.  For sources down to Hubble Ultra Deep Field (HUDF) depth ($\sim$30~AB), even a ten-meter class telescope would likely be sufficient to resolve the majority of the galaxies, assuming it is diffraction limited \citep[e.g.,][]{2006AJ....132.1729B}. Observers are helped here by the fact that the angular size of galaxies of a constant physical size does not decrease beyond $z\sim1.5$. Most of the cosmological information (see  \S\ref{sec:cosmo}) would come from these galaxies. However, the earliest bound collections of stars within a dark matter halos (the earliest `galaxies', see \S\ref{sec:gal_ev}) could be sufficiently small so as to remain unresolved even with large diameter telescopes ($D\sim100$m).
For the ``last galaxy survey", the observations would  be background limited for these faintest, most distant sources using a telescope in the vicinity of the Earth, greatly simplifying Eq.~\ref{inverse_snr}:

\begin{equation} \label{simple_inverse_snr}
    t =  \frac{\mathrm{SNR}^{2} S_{s}n_{p}}{S_{o}^{2}} .
\end{equation}
\noindent As we discuss in \S~\ref{sec:background}, for innovative choices of the location of the telescope(s) that might perform the last galaxy survey and sufficiently large diffraction limited telescopes, the assumption  that the sky background dominates source counts may start to break down, making Equation~\ref{simple_inverse_snr} and the following equations approximate.

Note that in  Equation~\ref{simple_inverse_snr} above and hereafter, we also make the significant simplifying assumption that detector noise (read noise and dark current) is subdominant to the other terms in Equation~\ref{inverse_snr} and that the detectors have $Q=1$. We discuss this assumption further in \S\ref{sec:focalplane}  As long as we consider point sources, the value of $S_{s}$ scales with $D^{2}$, while the size of the point source at the detector  over which the measurement is being made scales with $D^{-2}$, such that the sky count rate in photons/sec ($S_{s}n_{p}$ in the SNR equation) is a constant that depends on the sky brightness (flux per area), which we will call $f_{s}$. Because the photons/sec from the object $S_{o}$ scales like $D^{2}$, we have a sensitivity to point sources proportional to $D^{-4}$:
\begin{equation} \label{simple_inverse_snr_pt}
  t \propto \frac{\mathrm{SNR}^{2} f_{s}} {f_{o}^{2} D^{4}},
\end{equation}

\noindent where $f_{0}$ is the source flux. As mentioned above, most cosmological interesting sources at faint magnitudes and  high redshift would not present as point sources to diffraction limited telescopes with sufficiently large aperture. In that case, the total sky background under the source continues to scale as $D^{2}$, as the object does not shrink on the detector with increasing aperture. Thus, for extended sources, we get a sensitivity that scales as $D^{-2}$:

\begin{equation} \label{simple_inverse_snr2_ext}
  t \propto \frac{\mathrm{SNR}^{2} f_{s} A_{o}} {f_{o}^{2} D^{2}}.
\end{equation}

\noindent Here, $A_{o}$ is the area (e.g., arcsec$^{2}$) subtended by the source on the sky.  The other relevant terms for the time it would take for a given telescope to survey the sky to the required depth are the survey efficiency $\epsilon$, which we define as the fraction of time spent integrating as opposed to slewing/settling etc., as well as the focal plane area $A_{f}$ and the survey area $A_{s}$. Therefore, multiply Equation~\ref{simple_inverse_snr} by  $ A_{s}/(\epsilon{A_{f}})$ to get the total survey time using $N$ telescopes:

\begin{equation} \label{survey_time_sky_limit_pt}
\begin{split}
    t &= \left(  \frac{1}{3.15\times10^{7}} \right) \left(  \lambda (\mu\mathrm{m})\times10^{-4}\times206265 \right)^{2} \left( \frac{1240\times1.6\times10^{-12}}{\lambda(\mu\mathrm{m})\times10^{3}} \right) \left( \frac{1}{\mathrm{bandpass(Hz)}} \right)   \left( \frac{A_s}{\epsilon A_{f}} \right) \frac{\mathrm{SNR}^{2} f_{s}} {N f_{o}^{2} D^{4}} \\ &= C_{unres}\left( \frac{A_s}{\epsilon A_{f}} \right) \frac{\mathrm{SNR}^{2} f_{s}} {N f_{o}^{2} D^{4}}.
    \end{split}
\end{equation}

\noindent Here, $C_{\textrm{unres}}$ is a constant holding the conversion factors from the source and sky fluxes to photon counts per second for the unresolved sources, as well as the other factors such that the units of the equation are in years. Choosing an observation wavelength $\lambda=2\mu$m and a bandpass of $0.5\mu$m, we obtain a numerical value $C_{\textrm{unres}}=1.4\times10^{-30}$ year s$^-1$ erg Hz$^{-1}$ cm$^2$ arcsec$^2$.
 The $\lambda(\mu\mathrm{m})$ term (second from left) comes from the size of the PSF on the detector, which scales as $\lambda^{2}/D^{2}$ (hence the $D^{4}$ sensitivity scaling).

The full equation is similar for the case of extended (resolved) sources, but the galaxy size enters the equation:
\begin{equation}
\label{survey_time_sky_limit_ext}
\begin{split}
 t &=  \left(  \frac{1}{3.15\times10^{7}} \right) \left( \frac{1240\times1.6\times10^{-12}}{\lambda(\mu\mathrm{m})\times10^{3}} \right) \left( \frac{1}{\mathrm{bandpass(Hz)}} \right)   \left( \frac{A_s }{\epsilon A_{f}} \right) \frac{\mathrm{SNR}^{2} f_{s}A_{o}} {N f_{o}^{2} D^{2}} \\ &= C_{res} \left( \frac{A_s}{\epsilon A_{f}} \right) \frac{\mathrm{SNR}^{2} f_{s}A_{o}} {Nf_{o}^{2} D^{2}}.
 \end{split}
\end{equation}

\noindent Here,  $C_{\textrm{res}}=8.3\times10^{-34}$  year s$^{-1}$ erg Hz${^-1}$ is a constant holding the conversion factors for resolved sources and $A_{o}$ is the size of the extended object on the detector (arcsec$^2$). Obviously the time is reduced dramatically with a larger $D$, and we explore the implications of this in \S\ref{sec:aperture}. We also note that the sensitivity scales linearly with the sky background. This will lead us to consider novel ideas for reducing that background, as discussed in \S\ref{sec:background}.

\begin{deluxetable}{ccc}
\tabletypesize{\large}
%\rotate
\tablecaption{Values in the survey efficiency equations. \label{tab:quantities}}
\tablewidth{0pt}
\tablehead{
%\multicolumn{Field}{c}{$l$} & \multicolumn{7}{c}{RA DEC Area} \\[-0.2cm]
%\multicolumn{1}{c}{------------------} & \multicolumn{7}{c}{---------------------------------------------------------------------------------------------------------------------------} \\[-0.2cm]
\colhead{Quantity} & \colhead{Description} &
\colhead{Unit} \\
}
\startdata
$S_{o}$ & Source count rate at detector & photons per second \\
$S_{s}$ & Sky count rate per pixel at detector  & photons/second/pixel   \\
$f_{o}$ & Source flux & erg~sec$^{-1}$~cm$^-2$~Hz$^{-1}$ \\
%$f_{s}$ & Sky flux per area & MJy/sr \\
$f_{s}$ & Sky flux per area & erg~s$^{-1}$~cm$^{-2}$~Hz$^{-1}$~arcsec$^{-2}$  \\
$S_{d}$ & Dark Current & electron/second/pixel   \\
$R$ & Read Noise & electron/pixel   \\
D & Mirror Diameter & m   \\
$n_{p}$& number of pixels (diffraction limited) to cover object & -- \\
$A_{s}$ & Survey area & deg$^{2}$   \\
$A_{f}$ & Focal plane area & deg$^{2}$  \\
$A_{o}$ & Galaxy area on sky & arcsec$^{2}$  \\ 
Q  & quantum efficiency of telescope/detector system & --  \\
t  & time to complete survey & sec \\
N  & number of telescopes & -- \\
N$_{B}$ & number of wavelength channels & --\\
$C_{unres}$  & constant for unresolved observations (Eq.~\ref{survey_time_sky_limit_pt}) & year s$^-1$ erg Hz$^{-1}$ cm$^2$ arcsec$^2$  \\
$C_{res}$  & constant for resolved observations (Eq.~\ref{survey_time_sky_limit_ext}) & year s$^{-1}$ erg Hz${^-1}$ \\
%$C_{unres}$  & constant for unresolved observations  & year s$^-1$ erg Hz$^{-1}$ cm$^2$ arcsec$^2$  \\
%$C_{res}$  & constant for resolved observations  & year s$^{-1}$ erg Hz${^-1}$ \\
$\epsilon $ & Survey efficiency & -- 
\enddata
%    \begin{tablenotes}
    %\item[\textdagger] \textdagger \textit{Y} band obtained from CFHT-WIRCAM observations separate from the WIRDS survey.
%    \end{tablenotes}
\end{deluxetable}

\subsection{Redshifts and Spectroscopy} \label{sec:spec}
In this paper, we concentrate largely on the information needed for both cosmological and galaxy evolution studies. Above, we calculated the exposure time needed for a survey in a single NIR wavelength band.  However, it is likely that at least crude redshift, and thus spectroscopy, are required in the final galaxy survey.  Some might even suggest that spectroscopy should be the baseline for the final galaxy survey; here, however, we concentrate on a baseline photometric survey and in this section outline some crude scaling that would allow for spectroscopy.  The redshifts needed to extract and analyze the scientific content of the galaxies in the universe can range can range from fairly crude photometric redshifts requiring $\sim4$ passbands up to full slit spectra.   To fully probe the cosmological information contained in galaxy clustering (see \S\ref{sec:cosmo}), we estimate that redshifts with a scatter less than $1\%$ would be needed. It has been shown in other surveys (e.g., COSMOS, \citealp{Ilbert09}) that this level of redshift precision can be achieved with multiband photometry constituting a low-resolution spectrum, with an effective resolution $R\sim20$.  This would require an additional $\sim20$-$40$ passbands as described in the previous section using today's techniques.  Thus, the final galaxy survey might need many passbands, increasing the total survey time  by a factor roughly equal to the number of passbands (N$_{B}$) required if current techniques of single filters for each band are used.  However, as we discuss below in \S~\ref{sec:focalplane}, advances in detectors may mean that we can get wavelength and flux information from each pixel.  Redshifts from these photometric observations could be calibrated using an ultra-deep spectroscopic sample of a subset of galaxies spanning the observed space of galaxy properties, as described in %\citep{2015ApJ...813...53M}. 
 \cite{2015ApJ...813...53M}. With proper instrumentation, extremely deep spectra could be acquired in the very deep `galaxy evolution' final survey we describe in \S\ref{sec:gal_ev}.  Since the integration time to reach a given depth scales as the spectral resolution $R$ (as a result of reduced source flux per resolution element combined with the reduced sky background), significantly longer integration times would be needed with a slit spectrograph to achieve high resolution spectra. 
 High precision redshifts from $R\sim1000$ spectroscopy would require integration times  $\sim250$ times longer than  a fiducial single-band last galaxy survey (described below) with $R=\frac{\lambda}{\Delta\lambda}=\frac{2\mu m}{0.5 \mu m}=4$.
 While it is beyond the scope of this paper, it is clear that minimizing the number of passbands needed to extract redshift information from galaxies will play a crucial role in determining the time needed for the final galaxy survey.

\section{Final (?) Surveys}
\label{sec:surveys}
Here we describe three cases in which optical imaging and spatially unresolved spectroscopy would reach a natural `end'.

\subsection{Defining the information Content} \label{subsec:info}
The word ``information'' has a variety of meanings. Here, we use it to mean constraints on the parameters achievable from an optimal analysis of idealized data. In the sections that follow, we consider separately the parameters describing a minimal modification to the $\Lambda CDM$ cosmology and those that describe theoretical uncertainties on the formation and evolution of galaxies.

\subsection{Cosmology from Galaxies} \label{sec:cosmo}

Here, we assume that the combination of galaxy clustering (which provides information on Baryon Acoustic Oscillations and Redshift Space Distortions) and weak lensing measurements, using shapes, redshifts, and photometry, will be sufficient to break the degeneracy between galaxy bias and matter clustering. In the earliest epochs of structure formation in the universe, density fluctuations are small and the power spectrum of the distribution of matter will capture most of the available information in the density field.This power spectrum can be measured by using galaxies as tracers of mass (via galaxy clustering) or via measurements of the dark matter directly via weak gravitational lensing.  The binding requirements on a survey to extract all the available information from the power spectrum using galaxies as tracers will be set by the faintest tracers (galaxies) that occur in sufficient numbers to meaningfully constrain the matter power spectrum at the highest redshifts.

The noise in the galaxy power spectrum $P$ at some wavenumber is, to leading order, set by the shot noise and the cosmic variance:
\begin{equation}\label{eqn:pknoise}
\frac{\sigma_P}{P} = \sqrt{\frac{2}{N_k}}\left(1 + \frac{1}{\bar{n}P}\right)
\end{equation}

\noindent where the number of $k-$modes $N_k$ is set by the number of modes of wavelength $k$ sampled by the survey volume. With a single tracer population, there is clearly no point in pushing the tracer density $\bar{n}$ above $nP\sim 1$. With multiple tracers, however, the cosmic variance limit can be exceeded and increasing the tracer density continues to yield gains far past $n>1/P$ \citep{2009PhRvL.102b1302S,2009JCAP...10..007M}. As intensity mapping surveys of cold gas at these redshifts are likely to come to pass long before the last galaxy survey described here, it is reasonable to assume that the limits to cosmology will be set by the number of available galaxies, rather than cosmic variance. This means that we need to design a survey that will produce a comprehensive census of the bulk of the galaxies in the epoch where galaxy formation begins; for the purposes of this exercise, that means achieving completeness down to the limit where the galaxy luminosity function peaks.

The first galaxies are thought to form during the reionization of the universe via a few different channels, all of which are sensitive to the degree to which primordial gas is enriched by the first generation of stars, and the precise mechanisms by which it can cool to begin forming the stellar content of the first galaxies. Models for the first generation of galaxies suggest that there are two primary formation mechanisms: Molecular cooling of $H_2$ in so-called minihaloes, with virial temperatures low enough ($<10^4$K) to allow for the formation of molecular Hydrogen (at $z\sim 20$), and less efficient atomic cooling somewhat later (at $z\sim 10$) in halos with virial temperatures above this threshold.  Simulations and theoretical models exhibit a wide variety of predictions for the properties of these galaxies, ranging from the \citet{2020arXiv200304442Q} estimate that the peak of the luminosity function for these early systems is around $M_{abs, UV}\sim -6$ to $M_{abs}\sim -8$ during reionization to \citet{2019MNRAS.488.2202J} simulations finding rest-frame luminosity functions that flatten fainter than about $M_{abs} = -12$. 

The galaxy luminosity function at these redshifts is poorly constrained by current data (but of course will be much better known by the time of the final galaxy survey, and be completely determined by the final survey itself). Optical and NIR surveys \citep{2019ApJ...880...25B,2018MNRAS.479.5184A} using HST or deep images from Subaru constrain the luminosity function to $M_{\rm abs, UV}$ $<$ -18, and the evidence for a turnover at the faint limit here is weak, at best \citep[see for example][]{2016ARA&A..54..761S}. The abundance of early star-forming galaxies does impact the rate at which the Universe reionizes and thus the optical depth to the CMB, $\tau$, which is well-constrained. Consistency with Planck determinations of $\tau$ appears to require that the rest frame UV luminosity function keep rising until $M_{UV}<-13$.

From the above range, we adopt as our fiducial targets for the cosmological survey the most optimistic case in both redshift and absolute magnitude, corresponding to galaxies with $M_{abs}= -12$ at $z=10$, which, for the case where the first sources are formed via atomic cooling as described above, captures the vast majority of the galaxies that exist in the Universe near the end of the epoch of reionization; this corresponds to a limiting AB apparent magnitude at $\sim2\mu m$ of $40.7$. These sources are expected to have physical sizes\footnote{At the extreme depths of the last galaxy survey, the notion of the physical size of these galaxies may be complicated by, for example, the detection of the circumgalactic medium (CGM). See, e.g.,  \citet{2020ApJ...896..125C} for recent developments on the understanding of the CGM.} of order $10$pc \citep{2019ARA&A..57..375S,2017arXiv171102090B}. Extrapolating current UV rest-frame  luminosity functions down to this limit typically yields source space densities of $1~{\rm Mpc}^{-3}$; we note that at this density, without multi-tracer gains, for a Planck 2015 \citep{2016A&A...594A..13P} cosmology, Equation~\ref{eqn:pknoise} indicates that the noise in the power spectrum becomes shot-noise dominated for $k>0.3~{\rm Mpc}^{-1}$, meaning that the tracer density and not cosmic variance limits the cosmological constraining power on scales below $20~{\rm Mpc}$.

The flux of these objects at $2\mu$m, $2\times10^{-36}{\rm erg~cm^{-2}~s^{-1}{Hz^{-1}}}$, is $10^{7}$ times fainter than the typical zodiacal background near Earth in a square arcsecond. In our calculations we use a fiducial value of the zodiacal background at 2$\mu$m as measured near Earth (e.g. L2) of $0.1$ MJy sr$^{-1}= 1.6 \times 10^{-29}$ erg~s$^{-1}$~cm$^{-2}$~Hz$^{-1}$~arcsec$^{-2}$ \citep{2000ApJ...536..550G}.
Of course the zodiacal background actually varies significantly over the sky, so, as with everything in this paper, these are only order of magnitude estimates of the numbers for a final galaxy survey. Evaluation of equation~\ref{inverse_snr} shows that, even with an ideal ($Q=1, R=0$, $\epsilon=1$) system, detecting such a source {\it with only zodiacal light backgrounds} would require a month of integration per band on a $50$m telescope. To get a sense for what is required to achieve a complete census of our fiducial $M_{\rm abs,UV}=-12$ targets, we allow a total survey time of $10$ years and assume we can achieve an optical design capable of a 20 ${\rm deg^2}$ field of view. Coverage of $20,000$ ${\rm deg^2}$ (essentially the entire extragalactic sky, as would be appropriate for the final galaxy survey) in the allotted time allows $3.2$ hours total integration time per field, which in turn requires a $280$m diameter primary collecting area. We note that this is broadly consistent with recent work \citep{2020arXiv200702946S} that claimed a $\sim100m$ diameter telescope would be needed to detect the first luminous objects in the universe, Pop III stars at apparent magnitude (AB) of $\sim39$.
At $z\sim10$, even  a 10pc source (on the low end of the size estimates for the galaxies we would use for the final cosmology survey) observed at 2$\mu$m  is resolved by a telescope of size D$\sim180$m.
Thus, writing survey and telescope parameters in terms of the fiducial survey parameters described above, equation~\ref{inverse_snr} simplifies to
% at z= 10 Wright gives 4.255 kpc/"
%4255pc per arcec so 10 pc=0.00235 arcec
% area= pi *(0.00235/2)^2~ 5x10^-6 arcsec^2
%to get required D for resolution, lambda/D=0.00235"
%D=lambda/0.00235"=(2x10^-6m/0.00235" )(3600"/deg)(180/pi)= 175
% at z= 15 Wright gives 3.176 kpc/"
%3176pc per arcec so 10 pc=0.003 arcec
% area= pi *(0.00235/2)^2~ 8x10^-6 arcsec^2

\begin{equation}\label{fiducial} 
\begin{split} 
 T_{\textrm{cosmo}} ({\rm years})= 10~{\rm years}
 \left(\frac{1}{N}\right)
 \left(\frac{N_B}{1}\right)
   \left(\frac{A_{o}}{5\times10^{-6}~\textrm{arcsec}^2}\right)
 \left(\frac{A_s}{20,000 ~\textrm{deg}^2}\right)
 \left(\frac{1}{\epsilon}\right)
 \left(\frac{280{\rm m}}{D}\right)^2 
\times \\
 \left(\frac{20~\textrm {sq deg}}{A_f}\right)
 \left(\frac{\rm SNR}{10}\right)^2
\left(\frac{f_s}{1.6\times 10^{-29} {\rm erg}\: {\rm s}^{-1} {\rm cm}^{-2}{\rm Hz}^{-1}{\rm arcsec}^{-2} }\right)
\left(\frac{     2\times   10^{-36} {\rm erg}\: {\rm s}^{-1} {\rm cm}^{-2}{\rm Hz}^{-1}}{f_o}\right)^2.
\end{split}
\end{equation}

\begin{figure}
    \centering
    \includegraphics[width=17cm]{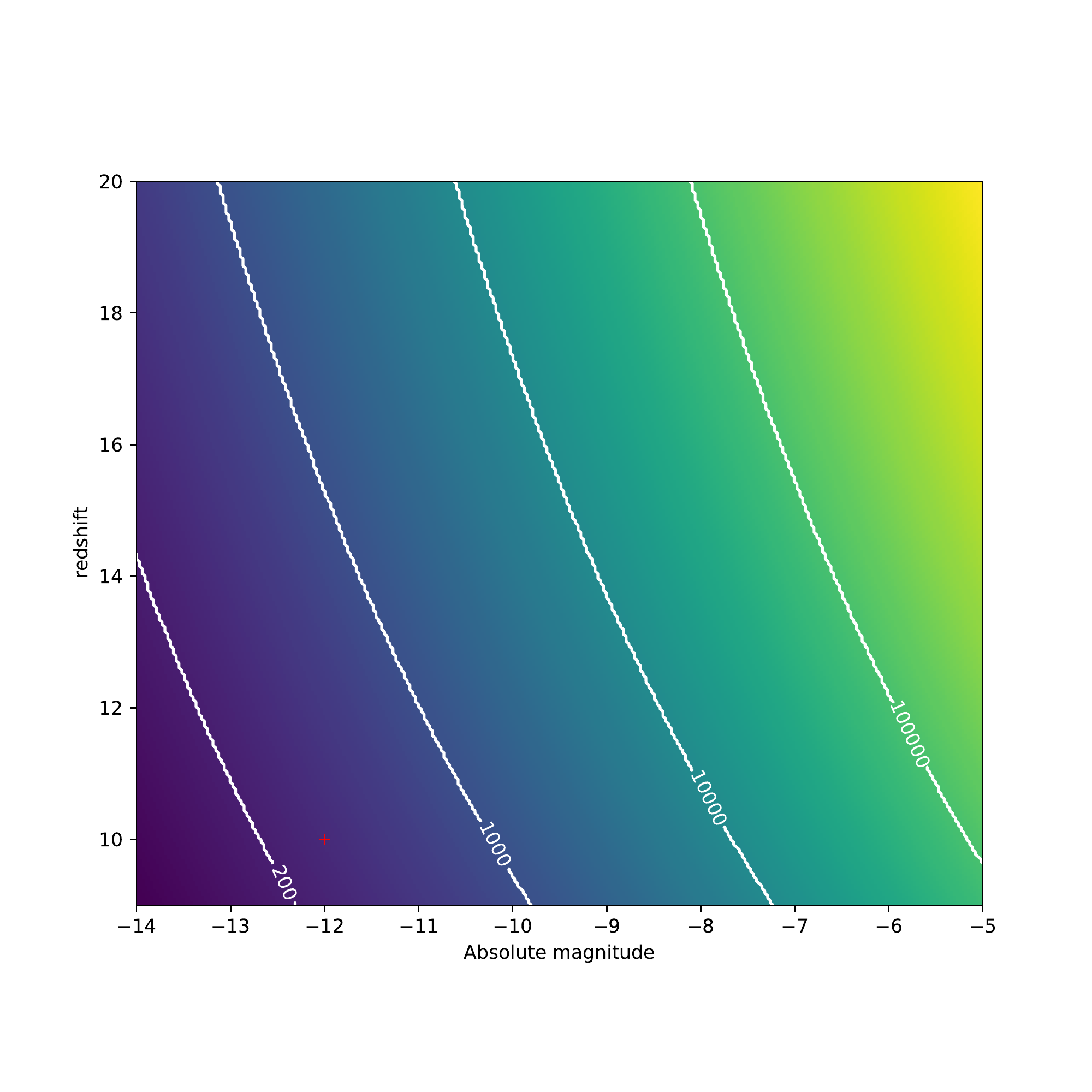}
    \caption{Required telescope aperture to complete a single-band, all-extragalactic-sky survey down to a given absolute magnitude and redshift in ten years. This assumes a 20 ${\rm deg}^2$ field of view, and a characteristic galaxy size of $10$pc. This plot is for typical background levels at L2; as we discuss below, background levels $\sim50$ times lower may be achievable elsewhere in the solar system and thus the numbers here could be reduced by a factor of $\sim \sqrt{50}$.  The fiducial cosmology survey as in equation~\ref{fiducial} is marked with a red cross. This figure shows the very wide range of possible collecting areas that the final galaxy survey might need depending on the properties of the universe that Nature provides.   These properties will become better known in the coming decades (starting with JWST next year) as we build toward the capability needed for the final galaxy survey.  
    }
   \label{fig:tradeoff2}
\end{figure}

If we have underestimated the sizes of the galaxies that contain cosmological information at these redshifts, and the true size is closer to 100 pc, the size $A_o$ goes up accordingly, requiring a much larger mirror ($\sim2$km!) due to the lower surface brightness of the galaxies (for a fixed intrinsic brightness).  Similarly, for the most challenging scenario in the range outlined above, where the luminosity function peaks at $M_{\rm abs, UV} = -6 $ galaxies forming in minihaloes at $z=20$, the apparent magnitude would be 49.1, with a source flux of $f_s=8\times10^{-40}{\rm erg}{\rm s}^{-1}{\rm cm}^{-2}{\rm Hz}^{-1}$. The required diameter increases proportionately, and the mirror size required to achieve a comprehensive census in the allotted time grows to a monstrous $50$km! Figure~\ref{fig:tradeoff2} shows a contour plot of the telescope size required for the final galaxy cosmology survey as a function of the proprieties (absolute magnitude and redshift of formation) of the earliest, faintest galaxies in the survey.   It should also be noted that the above calculations establish a requirement for single-band photometry; we do not here speculate on the photometric multiplexing capabilities of future instruments, which could mitigate the substantially stronger requirements associated with even low-resolution spectroscopy.

So far, we have assumed that the usable galaxy number density is limited only by the sensitivity and resolution of the survey. In modern surveys, however, a substantial fraction of the sources is typically blended with other nearby galaxies, reducing number of galaxies where shapes or photometry can be measured accurately enough for clustering or lensing measurements. For example, for the Vera Rubin Telescope's Legacy Survey of Space and Time, with an r-band limiting magnitude of $\sim27.5$, roughly $1/3$ of the galaxy population is expected to be blended with other nearby sources \citep{LSSTNeff}. Here, while the source density is astronomically higher than that available to existing programs, the corresponding improvement in angular resolution means that the bulk of the additional population consists of sources with physical sizes of $10-100$pc. At our fiducial physical source densities ($1 {\rm Mpc}^{-3}$, comoving), assuming every source is resolved, the entirety of the galaxy sample ($\sim5\times10^{12}$ galaxies, $0.02''$ typical radius) covers about $1\%$ of the sky. Despite this, source clustering will result in some significant blending between sources at similar redshifts. Given that the state of the art in deblending algorithms has advanced quickly over the last decade (see, for example, \citealt{2018A&C....24..129M}), however, it seems possible that much additional algorithmic progress will be made before a survey like that described here comes to pass. For this reason, we do not incorporate limitations due to blending in our forecasts.

It is clear that to extract all of the information about cosmology encoded in shapes and positions of galaxies would require an enormous telescope. This would likely be a key technical and cost driver of the final galaxy survey. Equation~\ref{fiducial} is useful in that it shows us where the lever arms are on technology and survey parameters to configure a telescope (or a fleet of N telescopes, as the galaxies in question are resolved for $D<200$m) to perform the last galaxy survey. While the universe will give us $f_o$ and extracting all of the information might require maximizing $A_s$, other aspects of this equation can be traded against each other.  Below, we explore some of these trade-offs, including mirror size $D$, focal plane size $A_f$, and novel orbits for reducing $f_s$ (a factor that could conceivably lower the required survey time $T$ by up to two orders of magnitude).

\subsection{Galaxy Evolution} \label{sec:gal_ev}

The goals of galaxy formation and evolution studies are constantly changing as new information is provided by progressively wider and deeper surveys.  
It is uncertain what the major open questions will be when we are at the technological point of considering having a broad band photometric measurement of virtually every kind of galaxy in the universe. 
Some questions one might envision today include what was the luminosity function and color distribution of the first galaxies and how did these distributions vary with environment.
To achieve these possible `end goals' we would need to detect and measure the luminosities and colors of the first, faintest galaxies that formed.
Here we consider what the hypothetical, `most challenging' class of galaxies might look like, and what resources would be needed to study them.

A galaxy is generally defined as a collection of stars which is internally bound to a dark matter halo.  Galaxies with only a few hundred stars have been identified in the Local Group. These galaxies appear to have formed the stars we observe today in an initial burst at high redshift ($z>10$) with rest-frame UV luminosities estimated to be $M_{UV} \sim -5$ and remained inactive since then \citep{Weisz++17}. Assuming a fiducial `first redshift' of $\sim 15$, these galaxies would have had an apparent magnitude of approximately $\sim 49$ at our fiducial observed wavelength of 2 $\mu$m.  This corresponds to a flux of $f_o=$  $8\times10^{-40}{\rm erg~cm^{2}~s^{-1}{Hz^{-1}}}$.
However, we note that these Local Group relics certainly do not represent the very first galaxies.  Ultra-faint Local Group dwarf galaxies have finite stellar metallicities, indicating  they must have formed in pre-enriched gas. The very initial burst of star formation which enriched the gas which the present day stars formed in likely had a different luminosity and spatial extent from the ancient surviving stars. Despite this, we select the Local Group ultra-faint dwarf galaxy analogs as our fiducial standard for the most challenging galaxies to observe for the purposes of this work. We do so because of the large theoretical uncertainties regarding galaxies formed of Population III stars and the lack of observational guidance. 
Of course, as we progress toward the final galaxy survey in the coming decades, our understanding of the most challenging galaxies to observe will likely evolve with our knowledge.

Typical studies which answer the sorts of galaxy formation questions outlined in the beginning of this section aim to measure the luminosity function in bins with uncertainties ranging from $\sim 1-10\%$ per bin. Uncertainties in galaxy luminosity function studies are driven by flux precision, the sample size, cosmic variance, and the ability to estimate observational completeness. 
Here, we assume that the observational completeness can be understood accurately in the case of a blank-field survey. Therefore, we consider here the survey area that would be necessary to obtain a sufficient sample size in order to obtain $\sim 1-10\%$ precision on the number of galaxies as a function of luminosity.  This would require bins of $\sim100$ galaxies for 10\% precision and bins with $10,000$ galaxies for 1\% precision.  

The spatial number densities of Ultra-Faint dwarf progenitors at high redshift is extremely uncertain. This in turn drives a large uncertainty in the survey area necessary to observe a sufficiently large number of objects to drive down Poisson uncertainty. Extrapolations based on the luminosity function of Local Group ultra-faint dwarf galaxies seem to indicate that the number density of the faintest objects may be as high as $\sim$100 per Mpc$^3$ at z$\sim 15$ \citep[see, e.g.,][]{2008ApJ...686..279K,2019ARA&A..57..375S}. In contrast, many theoretical studies predict a turnover in the UV luminosity function such that ultra-faint dwarf progenitors are actually relatively rare at high redshift. In this `pessimistic' scenario, the number density of these objects might be as low as 10$^{-2}$ per Mpc$^3$ \citep[see, e.g.,][]{2019MNRAS.490.2855Y}. Thus, the necessary survey area varies by 5 orders of magnitude between the optimistic and pessimistic cases for the number density of the earliest galaxies on the sky. At the highest potential surface densities of these galaxies, blending between the sources could become problematic for the analysis; it is beyond the scope of this paper to address the algorithms necessary for deblending such sources.

We assume here the more stringent goal of a 1\% uncertainty on the measurement of the luminosity function in the faintest bin, noting that the survey area would change significantly if the less stringent 10\% goal was adopted (a factor of 100 times smaller in survey area and thus a 100 times faster survey). 
We assume a redshift slice of $\delta z=0.1$ is required; this corresponds to a few Myr at $z=15$ and is sufficient because this is the lifetime of the shortest-lived stars and thus the fastest we could expect galaxies to evolve with redshift.
Thus, a given redshift slice at $z\sim15$ represents $\sim10$ Mpc in projection (co-moving). 
If the universe was completely uniform (without cosmic variance), given the angular diameter distance of $\sim$3 kpc/arcsecond at z$\sim 15$, we would need to observe 0.1 deg$^2$  in the optimistic case in which the faintest galaxies are abundant and $1,000$ deg$^2$ in the pessimistic case in which the faintest galaxies are rare, in order to obtain a volume containing sufficient numbers of objects to drive down Poisson uncertainties and extract the usable information encoded in the galaxies. To describe a fiducial last galaxy survey aimed at galaxy evolution, we could pick the mid-way (on a logarithmic scale) between the optimistic and pessimistic cases, setting the survey area at 10 deg$^2$. However, the above estimates do not take into account cosmic variance. 
When we calculate the cosmic variance contribution to the total galaxy number counts in slices of $\delta z=0.1$ over the fiducial survey area following \citep{2010ApJ...716L.229R}, we find that cosmic variance dominates the error in the counts for survey areas of order a square degree or smaller, pushing the required area to achieve $1\%$ precision to $\sim20~{\rm deg}^2$. Thus, we set this as the  fiducial area of our final galaxy evolution survey.
We again assume a  fiducial galaxy size of 10pc. These  galaxies have a larger apparent size at $z\sim15$ than they do for the cosmology survey above at $z\sim10$, and will therefore still be resolved by the large telescope that would be required for the final galaxy evolution survey. We thus calculate the time needed for a single band final galaxy evolution survey.  In reality, some aspects of galaxy evolution (e.g. star formation rates) we will need spectral coverage as described above in \S\ref{sec:spec}, meaning the numbers below will have to increase by a factor of $N_B \sim 5-20$ or we will have to have detectors that do wavelength multiplexing.  Thus, we can again put fiducial numbers into Equation~\ref{survey_time_sky_limit_ext} to give:

\begin{equation} \label{fiducial_res}
\begin{split}
 T_{\textrm{gal\_ev}} (years)=10
 \left(\frac{1}{N}\right)
  \left(\frac{N_B}{1}\right)
   \left(\frac{A_{o}}{8\times10^{-6}~\textrm{arcsec}^2}\right)
 \left(\frac{A_s}{20 ~\textrm{deg}^2}\right)
 \left(\frac{1}{\epsilon}\right)
 \left(\frac{13,000~\textrm{m}}{D}\right)^2
  \times \\
 \left(\frac{20~\textrm {sq deg}}{A_f}\right)
 \left(\frac{SNR}{10}\right)^2
 \left(\frac{f_s}{1.6\times 10^{-29} {\rm erg}\: {\rm s}^{-1} {\rm cm}^{-2}{\rm Hz}^{-1}{\rm arcsec}^{-2} }\right)
 \left(\frac{8\times10^{-40}{\rm erg}\:{\rm s}^{-1}{\rm cm}^{-2}{\rm Hz}^{-1}}{f_o}\right)^2.
\end{split}
\end{equation}

%https://latex.codecogs.com/legacy/eqneditor/editor.php to get equations into PDF

Thus, achieving single-band photometry of sufficient $M_{\rm abs,UV}=-5$ galaxies at $z=15$ would require ten years of integration on a $13$km-diameter telescope. Even readers who held a glimmer of hope in the previous section regarding the future feasibility of $\sim200$m space telescopes may be disheartened by this figure.

However, Nature has already shown us that it provided magnifying glasses (i.e. galaxy clusters) to study very high redshift objects.  Current programs that use massive galaxy clusters as lenses are able to survey around $500 {\rm Mpc}^3$ per cluster at five magnitudes of amplification \citep{2018MNRAS.479.5184A}. Several times $10^5$ such clusters exist above $10^{14}M_{\odot}$ at $z<0.7$.  The cosmology survey described above reaches a depth of 40.7 magnitudes. The galaxy evolution survey described above reaches a magnitude of 49;  with a factor of 5 magnitudes in magnification; this means that we must design a survey $49-5-40.7=3.2$ magnitudes deeper than the cosmology survey (or a factor of $\sim21$ greater in exposure time) to reach the required depth the final galaxy evolution survey. If we use the same 280m aperture, 20 deg$^2$ focal plane telescope as described in   \S~\ref{sec:cosmo}, this means we need about 67 hours per  pointing to reach the required depth. If each cluster magnifies a volume of $500 {\rm Mpc}^3$ and the number of 'first galaxies' is between $0.01$ and $100$ per Mpc$^3$ we need between $0.2$ and  $2,000$ clusters to reach the $10,000$ galaxies we estimate are needed above to complete the galaxy evolution survey.  If a single cluster suffices, we would need only 67 additional hours (one pointing) to complete the galaxy evolution survey.  If we assume that there are about 2 such clusters per square degree on the sky, each 20 square degree pointing would capture $\sim40$ clusters and we would need $50$ pointings to image $2,000$ clusters (our pessimistic estimate). This would take about 5 months on this telescope.  This, using the  same telescope described in \S~\ref{sec:cosmo} and leveraging the incredible power of gravitational lensing,  the final galaxy evolution survey might simply be a several day to several month perturbation on the cosmology survey.

\begin{figure}
    \centering
    \includegraphics{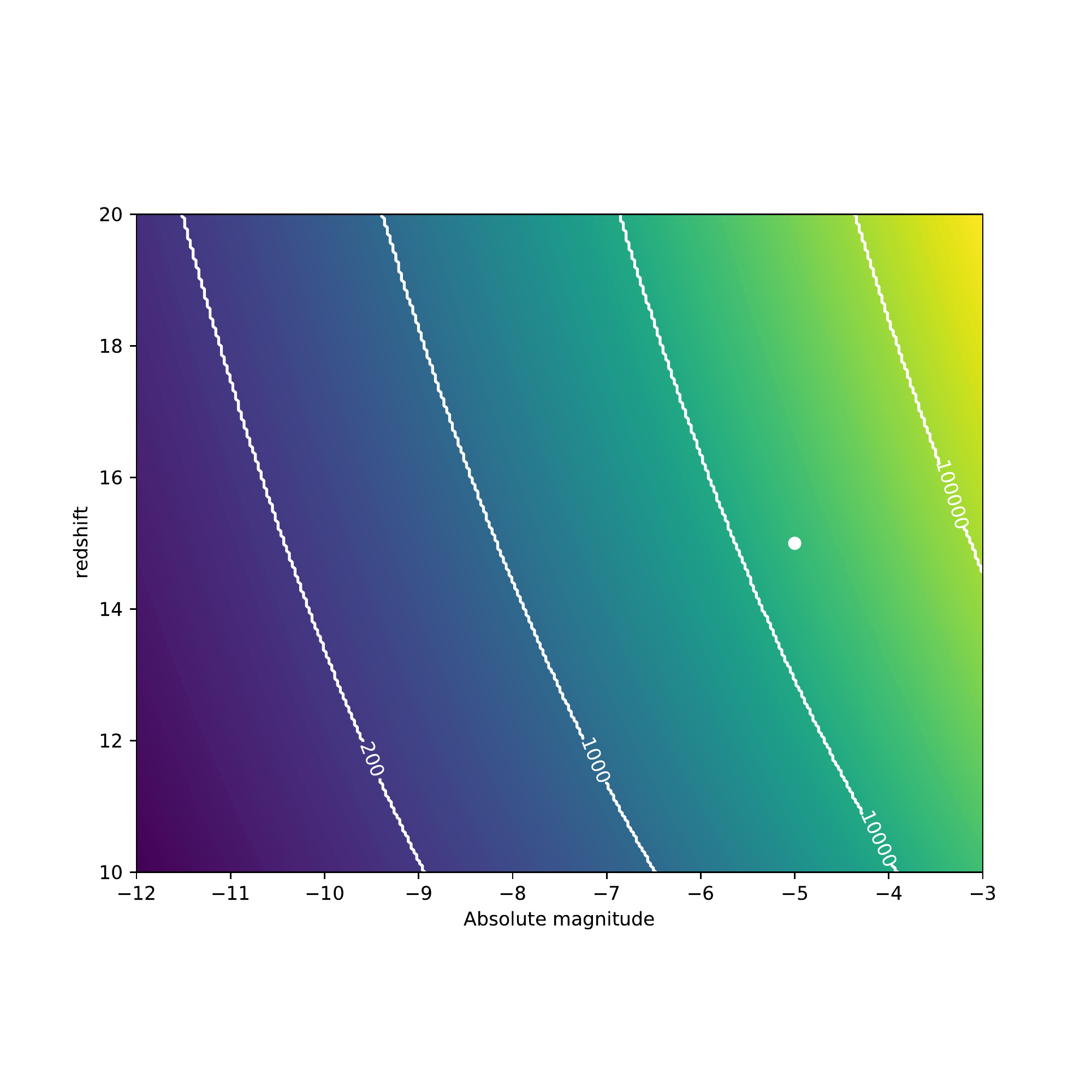}
    \caption{\textbf Required telescope aperture to complete the final galaxy evolution survey described in \S~\ref{sec:gal_ev} to a given absolute magnitude and redshift in ten years. This assumes a 20 ${\rm deg}^2$ field of view, and a characteristic galaxy size of $10$pc. This plot is for typical background levels at L2. This survey does not take into account the $\sim5$ magnitude gain due to gravitational lensing discussed in the text. The fiducial galaxy evolution survey as in equation~\ref{fiducial_res} is marked with a white dot.
    }
   \label{fig:tradeoff3}
\end{figure}

\subsection{Supernovae} \label{sec:SN}
Supernovae (SN) have long been a probe of cosmology and galaxy evolution, with these explosions briefly outshining their host galaxies.  Type Ia SN (SNIA) were used to discover the accelerating expansion of the universe in the 1990s \citep{1999ApJ...517..565P,1998AJ....116.1009R}.  An obvious ``end'' for any survey program involving SNIA would be to observe the earliest SNIA in the universe. The most massive SN1A progenitors are thought to be about 10\(M_\odot\). If we assume about 5 Myr for the mass transfer necessary to make the progenitor go SN, a 25 Myr lifetime for a $\sim10$\(M_\odot\) main sequence star, we could get a SN1AS as early as $\sim30$ Myr after the first PopII stars. The time it takes to form PopII stars is somewhat uncertain, with the lowest estimates about 220 Myr.  Thus, about 250 Myr after the Big Bang would be the earliest we could have a SN1A.  For a universe with  $H_{o}= 67.74$, $\Omega_{M}=0.309$, and 
$\Omega_{\Lambda}=0.691$, 
this corresponds to  $z\approx 16$ \citep{2006PASP..118.1711W}. At low redshifts, the light curve for SN1A spreads over $\sim50$ days. Thus, the $(1+z)$  time dilation means that we would need to measure the light curves of these earlier SN1A over the course of $2-3$ years rather than $\sim2$ months.   \citet{2006PASP..118.1711W} shows that the ratio of the luminosity distance $D_{L}$ for the aforementioned cosmology at $z=16$ to $z=1$ is a factor of $\sim25$.  Thus, to mimic the SN1A survey of Roman (for example), would require mirror diameter of 25 times that of Roman's $2.4m$ mirror; a $60$m space telescope observing at  $\sim20\mu m$ ($\sim1+z$ times the wavelength of Roman's survey) would be able to observe every unobscured SN1A in the universe. If the observations are done at NIR wavelengths ($\sim2\mu$ m), we may need an additional order of magnitude in either exposure time (or collecting area) to mimic the Roman survey due to lower flux in the rest frame UV  \citep[see, e.g.][]{2016MNRAS.461.1308F}.    SNIA rates have only been measured out to $z\approx2.5$ \citep{2014AJ....148...13R} and there is insufficient knowledge of the star formation density to even estimate rates out to the earliest possible SN1A at $z\sim16$.  Thus, any survey aimed at getting the earliest possible SN1A would require periodic observations of some pre-defined field (which could be up to the entire extragalactic sky, to piggyback on the other proposed surveys above at a cadence that would allow light curves to be measured over the course of several years).  

SNIA are expected to be much rarer than other types of SN at high redfshift. Therefore, a much better tracer of galaxies in the early universe would be core collapse SN (CCSN) or PopIII SN, as detailed in \citet{2006ApJ...637...80M}. Those authors calculated the rates of SN visibility for a hypothetical James Webb Space Telescope (JWST) survey extending to $z>16$. The rates of these types of SN at high redshift have much more robust predictions than SNIA, and \citet{2014MNRAS.442.1640D} provide predicted SN rates for various types of SN out to $z\sim25$. Numerous studies have shown that JWST is able to image these SN to even the faintest redfhifts, but is hampered by a small field of view and limited lifetime in the case of the highest redshifts, due to time dilation of the events \citep[see, for example,][]{2014MNRAS.442.1640D,2018MNRAS.479.2202H,2013MNRAS.436.1555D,2017arXiv171007005W}. A joint Roman/Subaru telescope survey over 5 years would discover a handful of SN at $z>6$ \citep{2019PASJ...71...59M}. However, sampling the light curves of SN at $z\sim20$ would require a survey that spanned a decade or more \citep{2018MNRAS.479.2202H}. The occurrence rates are estimated to be roughly constant with redshift over the range $10\leq z\leq 25$ (within an order of magnitude), with a $Mpc^3$ size box having 1 per million years to 1 per billion years, depending on the kind of SN \citep{2014MNRAS.442.1640D}.  Thus, an all-sky survey with a JWST or larger sized telescope could find tens of thousands of SN at $10<z<25$ (for example) each year. A desire to end SN surveys by viewing every visible SN in the universe  would thus drive survey duration (to a decade or more) and survey area.

\section{Mission Architecture} \label{sec:mission}
\subsection{Enabling Advances in Technology} \label{sec:tech}

Equation~\ref{survey_time_sky_limit_pt} illustrates the relevant factors in determining the time for a survey to reach the notional point source depth. Similarly, the relevant factors for extended sources are listed in Equation~\ref{survey_time_sky_limit_ext}. Which factors can we expect to work with the achieve the hugely ambitious survey outlined?

\subsection{Reducing background}\label{sec:background}

\begin{figure}
    \centering
    \includegraphics{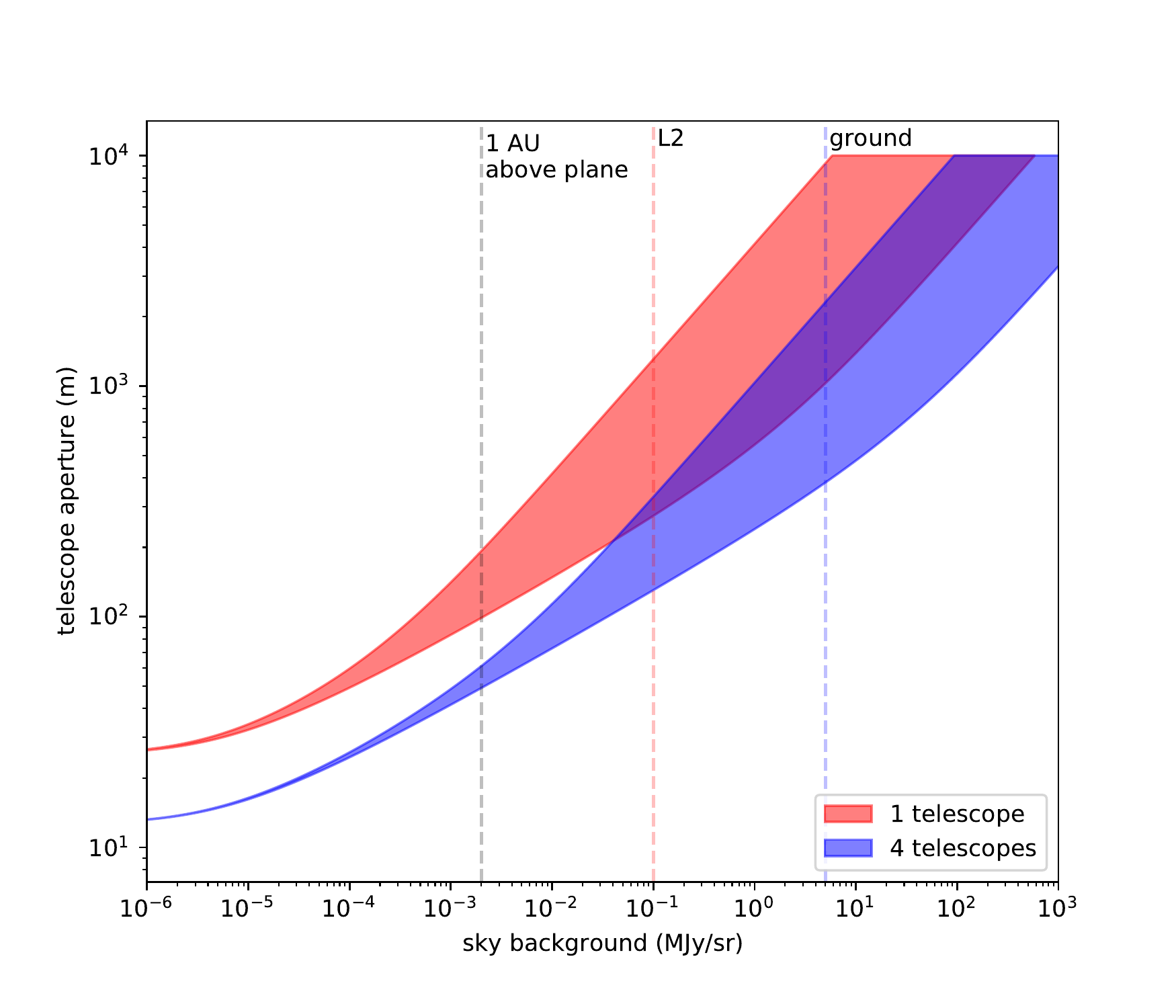}
    \caption{Required telescope aperture as a function of sky background needed to complete a single-band, all-extragalactic-sky survey down to a limiting AB magnitude of 40.7 in ten years. Since it is possible that a fleet of smaller telescope may be a more tractable path toward the last galaxy survey, we show this both for a single monolithic telescope and a fleet of telescopes (four in this example).    In order to resolve our fiducial smallest galaxies a $\sim180$m aperture is needed; thus at smaller apertures, multiple telescopes may not allow for resolving all the sources. This assumes a 20 ${\rm deg}^2$ field of view and other parameters take the fiducial values as in equation~\ref{fiducial}. The spread in the shaded areas show the effect of the as-yet unknown galaxy size of the faintest, earliest cosmology tracers; the lower end of the curves indicates 10 pc galaxies (our fiducial size) and the upper ends indicates galaxies that are 100 pc. The curve flattens below $\sim 10^{-5}$  MJy/sr because the angular size of the (resolved) sources is small enough that their photometric errors are dominated by source, rather than sky flux. However, it will likely not be possible to get below $10^{-3}$ MJy/sr in the solar system. Note that the vertical line indicating a ground-based telescope assumes diffraction limited seeing (and thus adaptive optics over the whole field of view) and assumes $\epsilon=1$, meaning the telescope would somehow have to observe day and night.
    }
   \label{fig:tradeoff1}
\end{figure}

Given that the time to complete a survey to a given depth and SNR scales with the sky background, one way to minimize survey time is to minimize sky background. Of course, much lower background is one of the key reason space telescopes are so effective.  The background in space\footnote{backgrounds for ground based telescopes are much larger, meaning that the final galaxy survey will not be feasible from the ground} at optical and NIR wavelengths, while much lower than that from the ground, is dominated by  zodiacal light emission from scattered dust in our solar system.     This zodiacal dust is an optically thin disk and thus the thermal emissivity and visible reflected flux scale with dust density.  This implies options to further reduce sky background for future telescopes beyond what is available to telescopes in Earth orbit or L2. \citet{2002ApJ...578.1009F} show that  at 2AU (in the plane of the Solar System) the zodiacal dust background is a factor $\sim5$ lower compared to the dust background at 1 AU.  Staying within the ecliptic plane but moving out beyond the asteroid belt (i.e. beyond $\sim4$ AU) would provide even lower zodiacal backgrounds \citep[up to a factor of $\sim10-100$;][]{2016Icar..264..369P}.  Significant reductions can also be achieved by going out of the ecliptic;  \citet{2002ApJ...578.1009F} show a factor of $\sim50$ reduction in the zodiacal dust by going $\sim1$AU out of the ecliptic plane.  Thus, a telescope outside of the ecliptic plane, possibly beyond the asteroid belt, would provide significantly lower sky background and thus faster, more sensitive surveys. Results from CIBER \citep{2017ApJ...839....7M} show that the Integrated Stellar Light (ISL) is an order of magnitude lower than the zodiacal light at $1-2\mu$m and the diffuse Galactic Light (DGL) is another order of magnitude lower than that (i.e., two orders of magnitude less than the zodiacal light).  Thus, factors of $50-100$ reduction in backgrounds are possible positioning by a telescope out of the ecliptic (and perhaps beyond the asteroid belt); it is beyond the scope of this paper to explore minimizing the DGL by putting a telescope outside of the Milky Way.  We show the relation between the telescope size needed for the final cosmology survey described in \S\ref{sec:cosmo} and the sky background (for ground, L2, and a factor of 50 below L2 levels) in Figure~\ref{fig:tradeoff1}.  

As we noted above, Equation~\ref{simple_inverse_snr} and the equations that follow assume that the observations for the final galaxy survey are background limited due to the very faint sources being observed.  However, Figure~\ref{fig:tradeoff1} shows that for the large, diffraction-limited telescopes being considered, this assumption starts to fail and the curves flatten out as the background gets lower.   For very small, but resolved (due to the large telescope diameter $D$) sources, the errors start to become dominated by source counts (which are concentrated in a few pixels), rather than background (which becomes lower for each pixel as $D$ increases and the pixel scale decreases correspondingly).  
Furthermore, the background numbers we use in this paper, like much of what we suppose, may be optimistic and work remains to be done to devise telescopes and observing strategies that would minimize these backgrounds (for example by excluding some region around stars to minimize ISL). Thus, for the final galaxy survey, there will be a tradeoff between telescope size D, number of telescopes , survey strategy, and telescope location (driving $f_s$) that goes beyond the level of detail considered in this paper.

\subsection{Extremely large focal plane}\label{sec:focalplane}

For a fixed focal plane area (as used in our nominal survey described in Equation~\ref{fiducial}), the physical size of the focal plane and number of pixels will be determined by the size $D$ of the telescope's primary mirror and the physical size of the pixels if we assume diffraction limited imaging.  The PSF size is given by the familiar $\frac{1.22\lambda}{D}$. Since we are trying to get estimates to an order of magnitude, we can simplify this to assume that each pixel must have an angular size of roughly $\frac{\lambda}{D}$. The linear size of the focal plane, as measured in pixels (where $N_{pix}$ is the total number of pixels) is thus
\begin{equation} \label{focal_plane_size}
  \sqrt{N_{pix}}=\frac{\sqrt{A_f}\frac{\pi}{180}}{\frac{\lambda}{D}}.
\end{equation}
Since both aperture size and focal plane size are typically large drivers of telescope cost, we consider here a scaling based on a fleet of  relatively modest-sized $10m$ telescopes (a fleet of such telescopes to finish the final galaxy surrey within the professional lifetime of an astronomer) \footnote{An aperture of $\sim180m$ is needed to resolve the smallest galaxies in the last galaxy survey; this could be achieved either with a mirror that size or  via   interferometry from smaller telescopes that achieve a $\sim180m$ baseline.  It is beyond the scope of this paper to explore this trade-off.} 
For a telescope with $D=10$m  (here we are  observing at 2$\mu m$ and a focal plane area of 20 square degrees, this would require 0.15 terapixels (150,000,000,000 pixels) This leads to a scaling relation where the number terapixels is given by
\begin{equation} \label{npix}
  \textrm{Number of Terapixels}=0.15\times\left(\frac{A_f}{20~\textrm{sq deg}}\right)\left(\frac{D}{10m}\right)^2\left(\frac{2\mu m}{\lambda}\right)^2.
\end{equation}
If we assume a physical pixel size of 5$\mu m$, then a $0.15$ terapixel focal plane would have a (relatively modest sounding) physical size of 3.8 square meters, or $\sim2$m on a side.  The linear dimension of the focal plane size scales linearly as the size of the pixels and as the square root of the number of pixels. Of course the telescope needed for the last galaxy survey as calculated above, would be much bigger if it is meant to resolve the smallest, faintest galaxies in the survey.  An aperture of D$=200m$ would require 60 terapixels.

We have assumed in our calculations that we have essentially perfect detectors.  This means detectors with 100\% quantum efficiency, zero read noise, and zero dark current. Note that for the survey in \S~\ref{sec:cosmo}, the telescopes only captures hundreds of photons for the faintest sources, so we do not worry about detector saturation here, even with very long exposures.   Furthermore, from Equation~\ref{snr}, where $S_{o}$ and $S_{s}$ depend on mirror diameter D, but $R$ and $S_{d}$ do not, detector noise will be subdominant at very large aperture sizes. Nonetheless, one area of technical development needed for the final galaxy survey is to push for enormous focal planes made of pixels with nearly perfect noise and efficiency properties. Focal planes of the unprecedented size being discussed here would entail engineering challenges related to power, cooling, and heat dissipation that are beyond the scope of this paper to address. Furthermore, we have also only calculated the survey times necessary  for a single band, whereas with today's techniques and detectors we would certainly have N$_B > 1$.  Advances in detectors such as Microwave Kinetic Inductance Detectors \citep[MKIDs; see, e.g., ][]{2017OExpr..2525894S} could allow for spectral as well as flux information to be acquired simultaneously, greatly increasing the spectral multiplexing ability of the telescope doing the last galaxy survey.

\subsection{Extremely large aperture}\label{sec:aperture}
    The sensitivity of a diffraction-limited telescope to unresolved sources scales as $D^{4}$, as shown in Equation~\ref{simple_inverse_snr_pt}. The rapid gain in sensitivity with aperture is due to a combination of the $D^{2}$ gain in collecting area and the ability to focus that light onto a smaller area of the detector (with area scaling like $D^{-2}$), thus dramatically reducing the sky background under the source. As long as the observations remain background-limited and the sources are unresolved, this scaling pertains. However, this scaling assumes that the faintest sources of interest remain point-like even at large apertures, and that the observations can remain background (rather than read noise) limited. Both of these assumptions break down at the aperture sizes discussed above.
 The desire to measure weak lensing shapes of galaxies pushes us to resolve those galaxies for the cosmology survey described above.  However, even if that were not a driver, we have shown that a single-band 10 year survey could require a mirror diameter sufficient to resolve our faintest, most distant sources. In the case of these resolved sources, the sensitivity scales as D$^2$. Thus, mirror diameter is a large driver of the power of the last galaxy survey no matter what resolution we desire. Given the dramatic scaling with diameter, aperture may dominate all other factors in the ability to reach a given source depth in a large survey in a given time. 
    
It is also a technical challenge to understand whether diffraction-limited telescope performance and the requisite optical stability could be achieved in space with extremely large apertures. Moreover, a very large aperture telescope would likely be a slow survey telescope due to the challenges of slewing, settling etc. This would likely  break our assumption of a telescope with high efficiency ($\epsilon \approx 1$). Multiple smaller aperture telescopes could reach the required depths, but an aperture of nearly 200m would be required to resolve 10 pc sources at $z>10$. Thus, a final survey would need to balance all of these factors - aperture, number of telescopes, survey efficiency, background level, and detector performance.

\subsection{Data Volume}\label{sec:data_vol}

The data volume for the telescope(s) taking the final galaxy survey will be enormous.  Assuming a $0.15$ terapixel focal plane (\S\ref{sec:focalplane}), 16 bits per pixel (the number for the next generation of H4RG detectors planned for Roman and a common number of astronomical detectors), each image would be 300GB. If we assume $100s$ exposures are taken continuously by the telescope, this would require a \textit{continuous} data transmission rate of $\sim$25Gbs in order to download all the data.  Of course, given the scaling relation in Equation~\ref{npix}, this could go up by a factor or $28^2\approx800$ for a mirror size that is 280m instead of 10m.  A 10 year survey with 100s exposures on a $0.15$ Tpix focal plane would correspond to nearly 1000PB of data.  NASA's Deep Space Optical Communications\footnote{https://www.nasa.gov/sites/default/files/atoms/files/fs\_dsoc\_factsheet\_150910.pdf} has plans to demonstrate 267Mbs from a distance of $0.5$AU. Thus, multiple orders of magnitude increase in optical communications data rate and several orders of magnitude increase in optical communications range would be needed to accommodate the final galaxy survey.

\section{Summary and Conclusions} \label{sec:conclusions}

We introduced a generic exposure time calculation for images of both resolved and unresolved faint astronomical sources. We used this equation, with some simplifying assumptions, to solve for the time needed to perform a survey reaching SNR$=10$ on objects of a certain flux with a given sky background.  This allowed us to show the scaling with telescope aperture and survey area.  We then identified three scientific scenarios in which a logical end of galaxy surveys could be defined.  For cosmology we posit that the information content of the universe would be saturated with a survey that achieves completeness down to the limit where the
galaxy luminosity function peaks.  For galaxy evolution, we posit that a survey of  $\sim10,000$ of the earliest  proto-galaxies(at their time of formation) that are analogs to conglomerations of $\sim100$ stars in our Local Group would suffice to answer extant questions in galaxy evolution.  For SN, we would like a survey that could see any SNIA or CCSN anywhere in the visible universe and follow that SN through its entire light curve. We showed that for the cosmology survey, with optimistic assumptions about the redshift and intrinsic brightness of the earliest useful sources, it would take a 20 deg$^2$ FOV and a 280m aperture to perform a single band survey in 10 years for the typical sky background a telescope at L2 sees.  If the luminosity function peaks at fainter galaxies or higher redshift, the required aperture could go up by orders of magnitude. Similar calculations for the galaxy evolution survey show that the much fainter galaxies required a much larger telescope (13 km!) for a smaller survey.  However, we showed that by making use of gravitational lensing magnification, we could answer the same galaxy evolution questions in a survey of a few days to a few months using the 280m cosmology survey telescope.  For SN, the requirements in aperture are relatively modest and the 280m telescope would easily accommodate a SN survey in conjunction with the cosmology survey above; time dilation of the light curves pushes for a survey time of more than a decade.

Having defined the  last galaxy survey (or surveys), we outlined the technological tall polls to achieving that survey.  One promising avenue to reducing the requirements for the last galaxy survey would be lowering the observed sky background, which could be done by going to a few AU from the Earth or an AU outside of the plane of the Solar System.  This would reduce the observed background by a factor of 50, enabling a factor of $\sim7$ reduction in mirror diameter.  We also identify that huge advances in telescope aperture, large focal planes with trillions of pixels, detector properties (including noise, QE, and wavelength multiplexing), and data transmission would be required to make even the most optimistic version of the last galaxy survey tractable.  However, all of these technical areas are ones that would have a myriad of benefits in astronomy and other space applications and we urge the community to continue the march toward extracting the available information content from galaxies in the universe.

\acknowledgements
\textbf{Acknowledgements:}
We thank the JPL Blue Skies Program run by  Leon Alkalai, the Blue Skies Program manager Adrian Stoica, and JPL's Chief Technology and Innovation Officer Tom Soderstrom for useful discussions and funding to pursue the ideas in this paper. This research was carried out at the Jet Propulsion Laboratory, California Institute of Technology, under a contract with the National Aeronautics and Space Administration (80NM0018D0004).   JPL is operated under contract for NASA by Caltech. We thank Steven Rodney for useful discussions about the earliest SN. We thank Bertrand Mennesson and Asantha Cooray for discussions about zodiacal backgrounds. Early discussions about ending galaxy surveys with David Schlegel helped formulate the ideas that led to this paper. We thank Andrew Benson and Kris Pardo for useful comments on the paper draft. JR thanks IPMU for a place to think and discuss the ideas that led to this paper. We thank the anonymous referee for enthusiasm about the subject matter and helpful suggestions to improve clarity. $\copyright$ 2020. All rights reserved

%% Note \ begin{thebibliography} is followed by an empty set of



\bibliography{end_gal.bib}

\begin{thebibliography}{}
\expandafter\ifx\csname natexlab\endcsname\relax\def\natexlab#1{#1}\fi
\providecommand{\url}[1]{\href{#1}{#1}}

\bibitem[{{Akeson} {et~al.}(2019){Akeson}, {Armus}, {Bachelet}, {Bailey},
  {Bartusek}, {Bellini}, {Benford}, {Bennett}, {Bhattacharya}, {Bohlin},
  {Boyer}, {Bozza}, {Bryden}, {Calchi Novati}, {Carpenter}, {Casertano},
  {Choi}, {Content}, {Dayal}, {Dressler}, {Dor{\'e}}, {Fall}, {Fan}, {Fang},
  {Filippenko}, {Finkelstein}, {Foley}, {Furlanetto}, {Kalirai}, {Gaudi},
  {Gilbert}, {Girard}, {Grady}, {Greene}, {Guhathakurta}, {Heinrich},
  {Hemmati}, {Hendel}, {Henderson}, {Henning}, {Hirata}, {Ho}, {Huff},
  {Hutter}, {Jansen}, {Jha}, {Johnson}, {Jones}, {Kasdin}, {Kelly}, {Kirshner},
  {Koekemoer}, {Kruk}, {Lewis}, {Macintosh}, {Madau}, {Malhotra}, {Mand el},
  {Massara}, {Masters}, {McEnery}, {McQuinn}, {Melchior}, {Melton},
  {Mennesson}, {Peeples}, {Penny}, {Perlmutter}, {Pisani}, {Plazas}, {Poleski},
  {Postman}, {Ranc}, {Rauscher}, {Rest}, {Roberge}, {Robertson}, {Rodney},
  {Rhoads}, {Rhodes}, {Ryan}, {Sahu}, {Sand}, {Scolnic}, {Seth}, {Shvartzvald},
  {Siellez}, {Smith}, {Spergel}, {Stassun}, {Street}, {Strolger}, {Szalay},
  {Trauger}, {Troxel}, {Turnbull}, {van der Marel}, {von der Linden}, {Wang},
  {Weinberg}, {Williams}, {Windhorst}, {Wollack}, {Wu}, {Yee}, \&
  {Zimmerman}}]{2019arXiv190205569A}
{Akeson}, R., {Armus}, L., {Bachelet}, E., {et~al.} 2019, arXiv e-prints,
  arXiv:1902.05569

\bibitem[{{Atek} {et~al.}(2018){Atek}, {Richard}, {Kneib}, \&
  {Schaerer}}]{2018MNRAS.479.5184A}
{Atek}, H., {Richard}, J., {Kneib}, J.-P., \& {Schaerer}, D. 2018, \mnras, 479,
  5184

\bibitem[{{Beckwith} {et~al.}(2006){Beckwith}, {Stiavelli}, {Koekemoer},
  {Caldwell}, {Ferguson}, {Hook}, {Lucas}, {Bergeron}, {Corbin}, {Jogee},
  {Panagia}, {Robberto}, {Royle}, {Somerville}, \&
  {Sosey}}]{2006AJ....132.1729B}
{Beckwith}, S. V.~W., {Stiavelli}, M., {Koekemoer}, A.~M., {et~al.} 2006, \aj,
  132, 1729

\bibitem[{{Bouwens} {et~al.}(2017){Bouwens}, {Illingworth}, {Oesch}, {Maseda},
  {Ribeiro}, {Stefanon}, \& {Lam}}]{2017arXiv171102090B}
{Bouwens}, R.~J., {Illingworth}, G.~D., {Oesch}, P.~A., {et~al.} 2017, arXiv
  e-prints, arXiv:1711.02090

\bibitem[{{Bouwens} {et~al.}(2019){Bouwens}, {Stefanon}, {Oesch},
  {Illingworth}, {Nanayakkara}, {Roberts-Borsani}, {Labb{\'e}}, \&
  {Smit}}]{2019ApJ...880...25B}
{Bouwens}, R.~J., {Stefanon}, M., {Oesch}, P.~A., {et~al.} 2019, \apj, 880, 25

\bibitem[{{Chang} {et~al.}(2013){Chang}, {Jarvis}, {Jain}, {Kahn}, {Kirkby},
  {Connolly}, {Krughoff}, {Peng}, \& {Peterson}}]{LSSTNeff}
{Chang}, C., {Jarvis}, M., {Jain}, B., {et~al.} 2013, \mnras, 434, 2121

\bibitem[{{Corlies} {et~al.}(2020){Corlies}, {Peeples}, {Tumlinson}, {O'Shea},
  {Lehner}, {Howk}, {O'Meara}, \& {Smith}}]{2020ApJ...896..125C}
{Corlies}, L., {Peeples}, M.~S., {Tumlinson}, J., {et~al.} 2020, \apj, 896, 125

\bibitem[{{de Souza} {et~al.}(2013){de Souza}, {Ishida}, {Johnson}, {Whalen},
  \& {Mesinger}}]{2013MNRAS.436.1555D}
{de Souza}, R.~S., {Ishida}, E.~E.~O., {Johnson}, J.~L., {Whalen}, D.~J., \&
  {Mesinger}, A. 2013, \mnras, 436, 1555

\bibitem[{{de Souza} {et~al.}(2014){de Souza}, {Ishida}, {Whalen}, {Johnson},
  \& {Ferrara}}]{2014MNRAS.442.1640D}
{de Souza}, R.~S., {Ishida}, E.~E.~O., {Whalen}, D.~J., {Johnson}, J.~L., \&
  {Ferrara}, A. 2014, \mnras, 442, 1640

\bibitem[{{DESI Collaboration} {et~al.}(2016){DESI Collaboration}, {Aghamousa},
  {Aguilar}, {Ahlen}, {Alam}, {Allen}, {Allende Prieto}, {Annis}, {Bailey},
  {Balland}, {Ballester}, {Baltay}, {Beaufore}, {Bebek}, {Beers}, {Bell},
  {Bernal}, {Besuner}, {Beutler}, {Blake}, {Bleuler}, {Blomqvist}, {Blum},
  {Bolton}, {Briceno}, {Brooks}, {Brownstein}, {Buckley-Geer}, {Burden},
  {Burtin}, {Busca}, {Cahn}, {Cai}, {Cardiel-Sas}, {Carlberg}, {Carton},
  {Casas}, {Castand er}, {Cervantes-Cota}, {Claybaugh}, {Close}, {Coker},
  {Cole}, {Comparat}, {Cooper}, {Cousinou}, {Crocce}, {Cuby}, {Cunningham},
  {Davis}, {Dawson}, {de la Macorra}, {De Vicente}, {Delubac}, {Derwent},
  {Dey}, {Dhungana}, {Ding}, {Doel}, {Duan}, {Ealet}, {Edelstein},
  {Eftekharzadeh}, {Eisenstein}, {Elliott}, {Escoffier}, {Evatt}, {Fagrelius},
  {Fan}, {Fanning}, {Farahi}, {Farihi}, {Favole}, {Feng}, {Fernandez},
  {Findlay}, {Finkbeiner}, {Fitzpatrick}, {Flaugher}, {Flender}, {Font-Ribera},
  {Forero-Romero}, {Fosalba}, {Frenk}, {Fumagalli}, {Gaensicke}, {Gallo},
  {Garcia-Bellido}, {Gaztanaga}, {Pietro Gentile Fusillo}, {Gerard},
  {Gershkovich}, {Giannantonio}, {Gillet}, {Gonzalez-de-Rivera},
  {Gonzalez-Perez}, {Gott}, {Graur}, {Gutierrez}, {Guy}, {Habib}, {Heetderks},
  {Heetderks}, {Heitmann}, {Hellwing}, {Herrera}, {Ho}, {Holland}, {Honscheid},
  {Huff}, {Hutchinson}, {Huterer}, {Hwang}, {Illa Laguna}, {Ishikawa},
  {Jacobs}, {Jeffrey}, {Jelinsky}, {Jennings}, {Jiang}, {Jimenez}, {Johnson},
  {Joyce}, {Jullo}, {Juneau}, {Kama}, {Karcher}, {Karkar}, {Kehoe}, {Kennamer},
  {Kent}, {Kilbinger}, {Kim}, {Kirkby}, {Kisner}, {Kitanidis}, {Kneib},
  {Koposov}, {Kovacs}, {Koyama}, {Kremin}, {Kron}, {Kronig}, {Kueter-Young},
  {Lacey}, {Lafever}, {Lahav}, {Lambert}, {Lampton}, {Land riau}, {Lang},
  {Lauer}, {Le Goff}, {Le Guillou}, {Le Van Suu}, {Lee}, {Lee}, {Leitner},
  {Lesser}, {Levi}, {L'Huillier}, {Li}, {Liang}, {Lin}, {Linder}, {Loebman},
  {Luki{\'c}}, {Ma}, {MacCrann}, {Magneville}, {Makarem}, {Manera}, {Manser},
  {Marshall}, {Martini}, {Massey}, {Matheson}, {McCauley}, {McDonald},
  {McGreer}, {Meisner}, {Metcalfe}, {Miller}, {Miquel}, {Moustakas}, {Myers},
  {Naik}, {Newman}, {Nichol}, {Nicola}, {Nicolati da Costa}, {Nie}, {Niz},
  {Norberg}, {Nord}, {Norman}, {Nugent}, {O'Brien}, {Oh}, {Olsen}, {Padilla},
  {Padmanabhan}, {Padmanabhan}, {Palanque-Delabrouille}, {Palmese},
  {Pappalardo}, {P{\^a}ris}, {Park}, {Patej}, {Peacock}, {Peiris}, {Peng},
  {Percival}, {Perruchot}, {Pieri}, {Pogge}, {Pollack}, {Poppett}, {Prada},
  {Prakash}, {Probst}, {Rabinowitz}, {Raichoor}, {Ree}, {Refregier}, {Regal},
  {Reid}, {Reil}, {Rezaie}, {Rockosi}, {Roe}, {Ronayette}, {Roodman}, {Ross},
  {Ross}, {Rossi}, {Rozo}, {Ruhlmann-Kleider}, {Rykoff}, {Sabiu}, {Samushia},
  {Sanchez}, {Sanchez}, {Schlegel}, {Schneider}, {Schubnell}, {Secroun},
  {Seljak}, {Seo}, {Serrano}, {Shafieloo}, {Shan}, {Sharples}, {Sholl},
  {Shourt}, {Silber}, {Silva}, {Sirk}, {Slosar}, {Smith}, {Smoot}, {Som},
  {Song}, {Sprayberry}, {Staten}, {Stefanik}, {Tarle}, {Sien Tie}, {Tinker},
  {Tojeiro}, {Valdes}, {Valenzuela}, {Valluri}, {Vargas-Magana}, {Verde},
  {Walker}, {Wang}, {Wang}, {Weaver}, {Weaverdyck}, {Wechsler}, {Weinberg},
  {White}, {Yang}, {Yeche}, {Zhang}, {Zhao}, {Zheng}, {Zhou}, {Zhou}, {Zhu},
  {Zou}, \& {Zu}}]{2016arXiv161100036D}
{DESI Collaboration}, {Aghamousa}, A., {Aguilar}, J., {et~al.} 2016, arXiv
  e-prints, arXiv:1611.00036

\bibitem[{{Fixsen} \& {Dwek}(2002)}]{2002ApJ...578.1009F}
{Fixsen}, D.~J., \& {Dwek}, E. 2002, \apj, 578, 1009

\bibitem[{{Foley} {et~al.}(2016){Foley}, {Pan}, {Brown}, {Filippenko}, {Fox},
  {Hillebrandt}, {Kirshner}, {Marion}, {Milne}, {Parrent}, {Pignata}, \&
  {Stritzinger}}]{2016MNRAS.461.1308F}
{Foley}, R.~J., {Pan}, Y.-C., {Brown}, P., {et~al.} 2016, \mnras, 461, 1308

\bibitem[{{Gorjian} {et~al.}(2000){Gorjian}, {Wright}, \&
  {Chary}}]{2000ApJ...536..550G}
{Gorjian}, V., {Wright}, E.~L., \& {Chary}, R.~R. 2000, \apj, 536, 550

\bibitem[{{Hartwig} {et~al.}(2018){Hartwig}, {Bromm}, \&
  {Loeb}}]{2018MNRAS.479.2202H}
{Hartwig}, T., {Bromm}, V., \& {Loeb}, A. 2018, \mnras, 479, 2202

\bibitem[{{Ilbert} {et~al.}(2009){Ilbert}, {Capak}, {Salvato}, {Aussel},
  {McCracken}, {Sanders}, {Scoville}, {Kartaltepe}, {Arnouts}, {Le Floc'h},
  {Mobasher}, {Taniguchi}, {Lamareille}, {Leauthaud}, {Sasaki}, {Thompson},
  {Zamojski}, {Zamorani}, {Bardelli}, {Bolzonella}, {Bongiorno}, {Brusa},
  {Caputi}, {Carollo}, {Contini}, {Cook}, {Coppa}, {Cucciati}, {de la Torre},
  {de Ravel}, {Franzetti}, {Garilli}, {Hasinger}, {Iovino}, {Kampczyk},
  {Kneib}, {Knobel}, {Kovac}, {Le Borgne}, {Le Brun}, {Le F{\`e}vre}, {Lilly},
  {Looper}, {Maier}, {Mainieri}, {Mellier}, {Mignoli}, {Murayama}, {Pell{\`o}},
  {Peng}, {P{\'e}rez-Montero}, {Renzini}, {Ricciardelli}, {Schiminovich},
  {Scodeggio}, {Shioya}, {Silverman}, {Surace}, {Tanaka}, {Tasca}, {Tresse},
  {Vergani}, \& {Zucca}}]{Ilbert09}
{Ilbert}, O., {Capak}, P., {Salvato}, M., {et~al.} 2009, \apj, 690, 1236

\bibitem[{{Jaacks} {et~al.}(2019){Jaacks}, {Finkelstein}, \&
  {Bromm}}]{2019MNRAS.488.2202J}
{Jaacks}, J., {Finkelstein}, S.~L., \& {Bromm}, V. 2019, \mnras, 488, 2202

\bibitem[{{Koposov} {et~al.}(2008){Koposov}, {Belokurov}, {Evans}, {Hewett},
  {Irwin}, {Gilmore}, {Zucker}, {Rix}, {Fellhauer}, {Bell}, \&
  {Glushkova}}]{2008ApJ...686..279K}
{Koposov}, S., {Belokurov}, V., {Evans}, N.~W., {et~al.} 2008, \apj, 686, 279

\bibitem[{{Laureijs} {et~al.}(2011){Laureijs}, {Amiaux}, {Arduini},
  {Augu{\`e}res}, {Brinchmann}, {Cole}, {Cropper}, {Dabin}, {Duvet}, {Ealet},
  {Garilli}, {Gondoin}, {Guzzo}, {Hoar}, {Hoekstra}, {Holmes}, {Kitching},
  {Maciaszek}, {Mellier}, {Pasian}, {Percival}, {Rhodes}, {Saavedra Criado},
  {Sauvage}, {Scaramella}, {Valenziano}, {Warren}, {Bender}, {Castander},
  {Cimatti}, {Le F{\`e}vre}, {Kurki-Suonio}, {Levi}, {Lilje}, {Meylan},
  {Nichol}, {Pedersen}, {Popa}, {Rebolo Lopez}, {Rix}, {Rottgering},
  {Zeilinger}, {Grupp}, {Hudelot}, {Massey}, {Meneghetti}, {Miller}, {Paltani},
  {Paulin-Henriksson}, {Pires}, {Saxton}, {Schrabback}, {Seidel}, {Walsh},
  {Aghanim}, {Amendola}, {Bartlett}, {Baccigalupi}, {Beaulieu}, {Benabed},
  {Cuby}, {Elbaz}, {Fosalba}, {Gavazzi}, {Helmi}, {Hook}, {Irwin}, {Kneib},
  {Kunz}, {Mannucci}, {Moscardini}, {Tao}, {Teyssier}, {Weller}, {Zamorani},
  {Zapatero Osorio}, {Boulade}, {Foumond}, {Di Giorgio}, {Guttridge}, {James},
  {Kemp}, {Martignac}, {Spencer}, {Walton}, {Bl{\"u}mchen}, {Bonoli},
  {Bortoletto}, {Cerna}, {Corcione}, {Fabron}, {Jahnke}, {Ligori}, {Madrid},
  {Martin}, {Morgante}, {Pamplona}, {Prieto}, {Riva}, {Toledo}, {Trifoglio},
  {Zerbi}, {Abdalla}, {Douspis}, {Grenet}, {Borgani}, {Bouwens}, {Courbin},
  {Delouis}, {Dubath}, {Fontana}, {Frailis}, {Grazian}, {Koppenh{\"o}fer},
  {Mansutti}, {Melchior}, {Mignoli}, {Mohr}, {Neissner}, {Noddle}, {Poncet},
  {Scodeggio}, {Serrano}, {Shane}, {Starck}, {Surace}, {Taylor},
  {Verdoes-Kleijn}, {Vuerli}, {Williams}, {Zacchei}, {Altieri}, {Escudero
  Sanz}, {Kohley}, {Oosterbroek}, {Astier}, {Bacon}, {Bardelli}, {Baugh},
  {Bellagamba}, {Benoist}, {Bianchi}, {Biviano}, {Branchini}, {Carbone},
  {Cardone}, {Clements}, {Colombi}, {Conselice}, {Cresci}, {Deacon}, {Dunlop},
  {Fedeli}, {Fontanot}, {Franzetti}, {Giocoli}, {Garcia-Bellido}, {Gow},
  {Heavens}, {Hewett}, {Heymans}, {Holland}, {Huang}, {Ilbert}, {Joachimi},
  {Jennins}, {Kerins}, {Kiessling}, {Kirk}, {Kotak}, {Krause}, {Lahav}, {van
  Leeuwen}, {Lesgourgues}, {Lombardi}, {Magliocchetti}, {Maguire}, {Majerotto},
  {Maoli}, {Marulli}, {Maurogordato}, {McCracken}, {McLure}, {Melchiorri},
  {Merson}, {Moresco}, {Nonino}, {Norberg}, {Peacock}, {Pello}, {Penny},
  {Pettorino}, {Di Porto}, {Pozzetti}, {Quercellini}, {Radovich}, {Rassat},
  {Roche}, {Ronayette}, {Rossetti}, {Sartoris}, {Schneider}, {Semboloni},
  {Serjeant}, {Simpson}, {Skordis}, {Smadja}, {Smartt}, {Spano}, {Spiro},
  {Sullivan}, {Tilquin}, {Trotta}, {Verde}, {Wang}, {Williger}, {Zhao},
  {Zoubian}, \& {Zucca}}]{2011arXiv1110.3193L}
{Laureijs}, R., {Amiaux}, J., {Arduini}, S., {et~al.} 2011, arXiv e-prints,
  arXiv:1110.3193

\bibitem[{{LSST Science Collaboration} {et~al.}(2009){LSST Science
  Collaboration}, {Abell}, {Allison}, {Anderson}, {Andrew}, {Angel}, {Armus},
  {Arnett}, {Asztalos}, {Axelrod}, \& et~al.}]{2009-Book-LSST}
{LSST Science Collaboration}, {Abell}, P.~A., {Allison}, J., {et~al.} 2009,
  ArXiv e-prints, arXiv:0912.0201

\bibitem[{{Masters} {et~al.}(2015){Masters}, {Capak}, {Stern}, {Ilbert},
  {Salvato}, {Schmidt}, {Longo}, {Rhodes}, {Paltani}, {Mobasher}, {Hoekstra},
  {Hildebrandt}, {Coupon}, {Steinhardt}, {Speagle}, {Faisst}, {Kalinich},
  {Brodwin}, {Brescia}, \& {Cavuoti}}]{2015ApJ...813...53M}
{Masters}, D., {Capak}, P., {Stern}, D., {et~al.} 2015, \apj, 813, 53

\bibitem[{{Matsuura} {et~al.}(2017){Matsuura}, {Arai}, {Bock}, {Cooray},
  {Korngut}, {Kim}, {Lee}, {Lee}, {Levenson}, {Matsumoto}, {Onishi},
  {Shirahata}, {Tsumura}, {Wada}, \& {Zemcov}}]{2017ApJ...839....7M}
{Matsuura}, S., {Arai}, T., {Bock}, J.~J., {et~al.} 2017, \apj, 839, 7

\bibitem[{{McDonald} \& {Seljak}(2009)}]{2009JCAP...10..007M}
{McDonald}, P., \& {Seljak}, U. 2009, \jcap, 2009, 007

\bibitem[{{Melchior} {et~al.}(2018){Melchior}, {Moolekamp}, {Jerdee},
  {Armstrong}, {Sun}, {Bosch}, \& {Lupton}}]{2018A&C....24..129M}
{Melchior}, P., {Moolekamp}, F., {Jerdee}, M., {et~al.} 2018, Astronomy and
  Computing, 24, 129

\bibitem[{{Mesinger} {et~al.}(2006){Mesinger}, {Johnson}, \&
  {Haiman}}]{2006ApJ...637...80M}
{Mesinger}, A., {Johnson}, B.~D., \& {Haiman}, Z. 2006, \apj, 637, 80

\bibitem[{{Moriya} {et~al.}(2019){Moriya}, {Wong}, {Koyama}, {Tanaka}, {Oguri},
  {Hilbert}, \& {Nomoto}}]{2019PASJ...71...59M}
{Moriya}, T.~J., {Wong}, K.~C., {Koyama}, Y., {et~al.} 2019, \pasj, 71, 59

\bibitem[{{Perlmutter} {et~al.}(1999){Perlmutter}, {Aldering}, {Goldhaber},
  {Knop}, {Nugent}, {Castro}, {Deustua}, {Fabbro}, {Goobar}, {Groom}, {Hook},
  {Kim}, {Kim}, {Lee}, {Nunes}, {Pain}, {Pennypacker}, {Quimby}, {Lidman},
  {Ellis}, {Irwin}, {McMahon}, {Ruiz-Lapuente}, {Walton}, {Schaefer}, {Boyle},
  {Filippenko}, {Matheson}, {Fruchter}, {Panagia}, {Newberg}, {Couch}, \&
  {Project}}]{1999ApJ...517..565P}
{Perlmutter}, S., {Aldering}, G., {Goldhaber}, G., {et~al.} 1999, \apj, 517,
  565

\bibitem[{{Planck Collaboration} {et~al.}(2016){Planck Collaboration}, {Ade},
  {Aghanim}, {Arnaud}, {Ashdown}, {Aumont}, {Baccigalupi}, {Banday},
  {Barreiro}, {Bartlett}, {Bartolo}, {Battaner}, {Battye}, {Benabed},
  {Beno{\^\i}t}, {Benoit-L{\'e}vy}, {Bernard}, {Bersanelli}, {Bielewicz},
  {Bock}, {Bonaldi}, {Bonavera}, {Bond}, {Borrill}, {Bouchet}, {Boulanger},
  {Bucher}, {Burigana}, {Butler}, {Calabrese}, {Cardoso}, {Catalano},
  {Challinor}, {Chamballu}, {Chary}, {Chiang}, {Chluba}, {Christensen},
  {Church}, {Clements}, {Colombi}, {Colombo}, {Combet}, {Coulais}, {Crill},
  {Curto}, {Cuttaia}, {Danese}, {Davies}, {Davis}, {de Bernardis}, {de Rosa},
  {de Zotti}, {Delabrouille}, {D{\'e}sert}, {Di Valentino}, {Dickinson},
  {Diego}, {Dolag}, {Dole}, {Donzelli}, {Dor{\'e}}, {Douspis}, {Ducout},
  {Dunkley}, {Dupac}, {Efstathiou}, {Elsner}, {En{\ss}lin}, {Eriksen},
  {Farhang}, {Fergusson}, {Finelli}, {Forni}, {Frailis}, {Fraisse},
  {Franceschi}, {Frejsel}, {Galeotta}, {Galli}, {Ganga}, {Gauthier}, {Gerbino},
  {Ghosh}, {Giard}, {Giraud-H{\'e}raud}, {Giusarma}, {Gjerl{\o}w},
  {Gonz{\'a}lez-Nuevo}, {G{\'o}rski}, {Gratton}, {Gregorio}, {Gruppuso},
  {Gudmundsson}, {Hamann}, {Hansen}, {Hanson}, {Harrison}, {Helou},
  {Henrot-Versill{\'e}}, {Hern{\'a}ndez-Monteagudo}, {Herranz}, {Hildebrand t},
  {Hivon}, {Hobson}, {Holmes}, {Hornstrup}, {Hovest}, {Huang}, {Huffenberger},
  {Hurier}, {Jaffe}, {Jaffe}, {Jones}, {Juvela}, {Keih{\"a}nen}, {Keskitalo},
  {Kisner}, {Kneissl}, {Knoche}, {Knox}, {Kunz}, {Kurki-Suonio}, {Lagache},
  {L{\"a}hteenm{\"a}ki}, {Lamarre}, {Lasenby}, {Lattanzi}, {Lawrence}, {Leahy},
  {Leonardi}, {Lesgourgues}, {Levrier}, {Lewis}, {Liguori}, {Lilje},
  {Linden-V{\o}rnle}, {L{\'o}pez-Caniego}, {Lubin}, {Mac{\'\i}as-P{\'e}rez},
  {Maggio}, {Maino}, {Mandolesi}, {Mangilli}, {Marchini}, {Maris}, {Martin},
  {Martinelli}, {Mart{\'\i}nez-Gonz{\'a}lez}, {Masi}, {Matarrese}, {McGehee},
  {Meinhold}, {Melchiorri}, {Melin}, {Mendes}, {Mennella}, {Migliaccio},
  {Millea}, {Mitra}, {Miville-Desch{\^e}nes}, {Moneti}, {Montier}, {Morgante},
  {Mortlock}, {Moss}, {Munshi}, {Murphy}, {Naselsky}, {Nati}, {Natoli},
  {Netterfield}, {N{\o}rgaard-Nielsen}, {Noviello}, {Novikov}, {Novikov},
  {Oxborrow}, {Paci}, {Pagano}, {Pajot}, {Paladini}, {Paoletti}, {Partridge},
  {Pasian}, {Patanchon}, {Pearson}, {Perdereau}, {Perotto}, {Perrotta},
  {Pettorino}, {Piacentini}, {Piat}, {Pierpaoli}, {Pietrobon}, {Plaszczynski},
  {Pointecouteau}, {Polenta}, {Popa}, {Pratt}, {Pr{\'e}zeau}, {Prunet},
  {Puget}, {Rachen}, {Reach}, {Rebolo}, {Reinecke}, {Remazeilles}, {Renault},
  {Renzi}, {Ristorcelli}, {Rocha}, {Rosset}, {Rossetti}, {Roudier},
  {Rouill{\'e} d'Orfeuil}, {Rowan-Robinson}, {Rubi{\~n}o-Mart{\'\i}n},
  {Rusholme}, {Said}, {Salvatelli}, {Salvati}, {Sandri}, {Santos},
  {Savelainen}, {Savini}, {Scott}, {Seiffert}, {Serra}, {Shellard}, {Spencer},
  {Spinelli}, {Stolyarov}, {Stompor}, {Sudiwala}, {Sunyaev}, {Sutton},
  {Suur-Uski}, {Sygnet}, {Tauber}, {Terenzi}, {Toffolatti}, {Tomasi},
  {Tristram}, {Trombetti}, {Tucci}, {Tuovinen}, {T{\"u}rler}, {Umana},
  {Valenziano}, {Valiviita}, {Van Tent}, {Vielva}, {Villa}, {Wade}, {Wandelt},
  {Wehus}, {White}, {White}, {Wilkinson}, {Yvon}, {Zacchei}, \&
  {Zonca}}]{2016A&A...594A..13P}
{Planck Collaboration}, {Ade}, P.~A.~R., {Aghanim}, N., {et~al.} 2016, \aap,
  594, A13

\bibitem[{{Poppe}(2016)}]{2016Icar..264..369P}
{Poppe}, A.~R. 2016, \icarus, 264, 369

\bibitem[{{Qin} {et~al.}(2020){Qin}, {Mesinger}, {Park}, {Greig}, \&
  {Munoz}}]{2020arXiv200304442Q}
{Qin}, Y., {Mesinger}, A., {Park}, J., {Greig}, B., \& {Munoz}, J.~B. 2020,
  arXiv e-prints, arXiv:2003.04442

\bibitem[{{Riess} {et~al.}(1998){Riess}, {Filippenko}, {Challis},
  {Clocchiatti}, {Diercks}, {Garnavich}, {Gilliland}, {Hogan}, {Jha},
  {Kirshner}, {Leibundgut}, {Phillips}, {Reiss}, {Schmidt}, {Schommer},
  {Smith}, {Spyromilio}, {Stubbs}, {Suntzeff}, \&
  {Tonry}}]{1998AJ....116.1009R}
{Riess}, A.~G., {Filippenko}, A.~V., {Challis}, P., {et~al.} 1998, \aj, 116,
  1009

\bibitem[{{Robertson}(2010)}]{2010ApJ...716L.229R}
{Robertson}, B.~E. 2010, \apjl, 716, L229

\bibitem[{{Rodney} {et~al.}(2014){Rodney}, {Riess}, {Strolger}, {Dahlen},
  {Graur}, {Casertano}, {Dickinson}, {Ferguson}, {Garnavich}, {Hayden}, {Jha},
  {Jones}, {Kirshner}, {Koekemoer}, {McCully}, {Mobasher}, {Patel}, {Weiner},
  {Cenko}, {Clubb}, {Cooper}, {Filippenko}, {Frederiksen}, {Hjorth},
  {Leibundgut}, {Matheson}, {Nayyeri}, {Penner}, {Trump}, {Silverman}, {U},
  {Azalee Bostroem}, {Challis}, {Rajan}, {Wolff}, {Faber}, {Grogin}, \&
  {Kocevski}}]{2014AJ....148...13R}
{Rodney}, S.~A., {Riess}, A.~G., {Strolger}, L.-G., {et~al.} 2014, \aj, 148, 13

\bibitem[{{Schauer} {et~al.}(2020){Schauer}, {Drory}, \&
  {Bromm}}]{2020arXiv200702946S}
{Schauer}, A. T.~P., {Drory}, N., \& {Bromm}, V. 2020, arXiv e-prints,
  arXiv:2007.02946

\bibitem[{{Seljak}(2009)}]{2009PhRvL.102b1302S}
{Seljak}, U. 2009, \prl, 102, 021302

\bibitem[{{Simon}(2019)}]{2019ARA&A..57..375S}
{Simon}, J.~D. 2019, \araa, 57, 375

\bibitem[{{Stark}(2016)}]{2016ARA&A..54..761S}
{Stark}, D.~P. 2016, \araa, 54, 761

\bibitem[{{Szypryt} {et~al.}(2017){Szypryt}, {Meeker}, {Coiffard}, {Fruitwala},
  {Bumble}, {Ulbricht}, {Walter}, {Daal}, {Bockstiegel}, {Collura}, {Zobrist},
  {Lipartito}, \& {Mazin}}]{2017OExpr..2525894S}
{Szypryt}, P., {Meeker}, S.~R., {Coiffard}, G., {et~al.} 2017, Optics Express,
  25, 25894

\bibitem[{{Wang} {et~al.}(2017){Wang}, {Baade}, {Baron}, {Bernard}, {Bromm},
  {Brown}, {Clayton}, {Cooke}, {Croton}, {Curtin}, {Drout}, {Doi}, {Dominguez},
  {Finkelstein}, {Gal-Yam}, {Geil}, {Heger}, {Hoeflich}, {Jian}, {Krisciunas},
  {Koekemoer}, {Lunnan}, {Maeda}, {Maund}, {Modjaz}, {Mould}, {Nomoto},
  {Nugent}, {Patat}, {Pacucci}, {Phillips}, {Rest}, {Regos}, {Sand}, {Sparks},
  {Spyromilio}, {Staveley-Smith}, {Suntzeff}, {Uddin}, {Villarroel}, {Vinko},
  {Whalen}, {Wheeler}, {Wood-Vasey}, {Yang}, \& {Yue}}]{2017arXiv171007005W}
{Wang}, L., {Baade}, D., {Baron}, E., {et~al.} 2017, arXiv e-prints,
  arXiv:1710.07005

\bibitem[{{Weisz} \& {Boylan-Kolchin}(2017)}]{Weisz++17}
{Weisz}, D.~R., \& {Boylan-Kolchin}, M. 2017, \mnras, 469, L83

\bibitem[{{Wright}(2006)}]{2006PASP..118.1711W}
{Wright}, E.~L. 2006, \pasp, 118, 1711

\bibitem[{{Yung} {et~al.}(2019){Yung}, {Somerville}, {Popping}, {Finkelstein},
  {Ferguson}, \& {Dav{\'e}}}]{2019MNRAS.490.2855Y}
{Yung}, L.~Y.~A., {Somerville}, R.~S., {Popping}, G., {et~al.} 2019, \mnras,
  490, 2855

\end{thebibliography}



\end{document}